\documentclass[%
reprint, superscriptaddress,
showpacs,
 amsmath,amssymb,
 aps,
]{revtex4-1}
\usepackage[svgnames]{xcolor}
\usepackage{lineno}

\usepackage{multirow,booktabs,footnote}
\usepackage{epsfig,float,subfigure,mathrsfs,mathtools,ulem,comment}
\definecolor{darkbrown}{rgb}{0.4, 0.26, 0.13}
\definecolor{darksienna}{rgb}{0.24, 0.08, 0.08}
\definecolor{darkpowderblue}{rgb}{0.0, 0.2, 0.6}

\usepackage[colorlinks]{hyperref}
\hypersetup{linkcolor={darkpowderblue},citecolor={cyan},urlcolor={darksienna}}  
\usepackage{orcidlink}
\usepackage{euscript,graphicx}

\usepackage{autobreak,cleveref}
\usepackage{amsthm,dsfont,amsfonts,amsmath,amssymb,wasysym}
\usepackage{euscript,color,fontenc,textcomp,relsize}
\usepackage{bm,url,float}

\makeatletter
\def\@hangfrom@section#1#2#3{\@hangfrom{#1#2}#3}
\def\@hangfroms@section#1#2{#1#2}
\makeatother

\newcommand{\gec}{\texttt{GWecc.jl}}
\newcommand{\epr}{\texttt{ENTERPRISE}~}
\newcommand{\gh}{\texttt{GW\_hyp}}
\newcommand{\ghn}{\texttt{GWhyp}~}
\newcommand{\nau}{\texttt{NAUTILUS}~}
\newcommand{\ptm}{\texttt{PTMCMCSampler}}

\DeclareMathAlphabet\mathbfcal{OMS}{cmsy}{b}{n}

\allowdisplaybreaks

\begin{document}

\title{Efficient prescription to search for linear gravitational wave memory from hyperbolic black hole encounters and its application to the NANOGrav 12.5-year dataset}


\author{Subhajit Dandapat \orcidlink{0000-0003-4965-9220}}

\email{subhajit.phy97@gmail.com}
\affiliation{Department of Astronomy and Astrophysics,
 Tata Institute of Fundamental Research, Mumbai 400005, Maharashtra, India}

\author{Abhimanyu Susobhanan \orcidlink{0000-0002-2820-0931}} 
\affiliation{Center for Gravitation Cosmology and Astrophysics, University of Wisconsin-Milwaukee, Milwaukee, WI 53211, USA}

\author{Lankeswar Dey \orcidlink{0000-0002-2554-0674}}
\affiliation{Department of Physics and Astronomy, West Virginia University, P.O. Box 6315, Morgantown, WV 26506, USA}
\affiliation{Center for Gravitational Waves and Cosmology, West Virginia University, Chestnut Ridge Research Building, Morgantown, WV 26505, USA}

\author{A.~Gopakumar \orcidlink{0000-0003-4274-4369}}
\affiliation{Department of Astronomy and Astrophysics,
 Tata Institute of Fundamental Research, Mumbai 400005, Maharashtra, India}

\author{Paul T.~Baker \orcidlink{0000-0003-2745-753X}}
\affiliation{Department of Physics and Astronomy, Widener University, One University Place, Chester, PA 19013, USA}

\author{Philippe Jetzer \orcidlink{0000-0001-6754-0296}}
\affiliation{Physik-Institut, University of Zurich, Winterthurerstrasse 190, 8057 Zurich, Switzerland}
 
\newcommand{\subh}[1]{{\color{magenta} \it{\textbf{Subhajit: #1}} }}

\newcommand{\blue}[1]{\textcolor{blue}{#1}}
\newcommand{\green}[1]{\textcolor[HTML]{1E4D2B}{#1}}
\begin{abstract}

Burst with memory events are potential transient gravitational wave sources for the maturing pulsar timing array (PTA) efforts.
We provide a computationally efficient prescription to model pulsar timing residuals induced by supermassive black hole pairs in general relativistic hyperbolic trajectories employing a Keplerian-type parametric solution.
Injection studies have been pursued on the resulting bursts with linear GW memory (LGWM) events with simulated datasets to test the performance of our pipeline, followed by its application to the publicly available NANOGrav 12.5-year (NG12.5) dataset.
Given the absence of any evidence of LGWM events within the real NG12.5 dataset, we impose 95\% upper limits on the PTA signal amplitude as a function of the sky location of the source and certain characteristic frequency ($n$) of the signal. 
The upper limits are computed using a signal model that takes into account the presence of intrinsic timing noise specific to each pulsar, as well as a common, spatially uncorrelated red noise, alongside the LGWM signal.
Our investigations reveal that the 95\% upper limits on LGWM amplitude, marginalized over all other parameters, is 3.48 $\pm 0.51 \ \mu$s for $n>3.16$ nHz.
This effort should be relevant for constraining both burst and memory events in the upcoming International Pulsar Timing Array data releases.



\end{abstract}

\maketitle

\section{Introduction}

Pulsar Timing Arrays (PTAs) are experiments that are capable of detecting gravitational waves (GWs) in the nanohertz frequency range by monitoring ensembles of millisecond pulsars (MSPs) \citep{sazhin1978pta,foster1990pta}.
A PTA operates by synthesizing a galaxy-sized GW detector out of its ensemble of MSPs acting as accurate celestial clocks.
There are six PTA collaborations active worldwide: the European PTA \citep[EPTA:][]{desvignes2016eptadr1}, the Indian PTA \citep[InPTA:][]{tarafdar2022inptadr1}, the Australia-based Parkes PTA \citep[PPTA:][]{manchester2013pptadr1}, the North American Nanohertz Observatory for Gravitational Waves \citep[NANOGrav:][]{demorest2013ng5yr}, the Chinese PTA \citep[CPTA:][]{xu2023cptagwb}, and the South Africa-based MeerKAT PTA \citep[MPTA:][]{miles2023mptadr1}.
The International Pulsar Timing Array \citep[IPTA:][]{verbiest2016iptadr1} consortium aims to advance the prospects of PTA science by combining data and resources from a subset of these regional PTA collaborations.

Recently, evidence for the presence of a stochastic GW background (GWB) was reported independently by NANOGrav \cite{agazie2023ng15yrgwb}, EPTA+InPTA \cite{antoniadis2023eptadr2gwb}, PPTA \cite{reardon2023pptadr3gwb}, and CPTA\cite{xu2023cptagwb} in their respective datasets. 
This signal manifests as a common-spectrum red noise process with Hellings-Downs cross-pulsar spatial correlations \cite{hellings1983upper}, with properties that are consistent across different datasets \cite{agazie2023ipta3p+}.
These exciting results have 
essentially inaugurated
the nanohertz window in GW astronomy, complementing the ground-based GW detectors operating in the 10 Hz-kHz frequency range \cite{aasi2015aligo,acernese2015avirgo,akutsu2020kagra}.
The source of the observed GWB signal remains inconclusive, with many astrophysical and cosmological processes such as a population of GW-emitting supermassive black hole binaries (SMBHBs), cosmic inflation, first-order phase transitions, cosmic strings, dark matter, etc. as possible candidates \cite{agazie2023ng15yrastro,afzal2023ng15yrnewphy,antoniadis2023eptadr2phys}.
Nevertheless, a population of inspiraling SMBHBs is considered to be the most prominent source of the GWB seen in these datasets \cite{burke-spolaor2019astrophysics}.
They were previously predicted to be detected first as a GWB, followed by the detection of individual sources resounding above the GWB \cite[e.g.][]{kelley2017prediction1,kelley2017prediction2,pol2021milestones,becsy2022exploring}.
Interestingly, both EPTA+InPTA and NANOGrav report the tentative presence of an individual SMBHB candidate signal with a GW frequency of $\sim 4$ nHz \cite{antoniadis2023second, agazie2023nanogravc}.
It is expected that the upcoming IPTA Data Release 3 will produce a higher-significance detection of the GWB and help characterize the astrophysical and/or cosmological processes producing the observed GWB \cite{agazie2023ipta3p+}.

PTAs in general are sensitive to three types of GW signals: (a) stochastic background (e.g., from an ensemble of SMBHBs), (b) persistent  GWs (e.g., from individual SMBHBs), and (c) burst events \cite{burke-spolaor2019astrophysics}.
Burst events to which PTAs are sensitive include cosmic string cusps \cite{damour2000cusp}, SMBHB mergers \cite{madison2014merger}, hyperbolic encounters of SMBH pairs \cite{dandapat2023gravitational}, near-field events such as binary mergers and supernovae in our Galaxy \cite{madison2017memory}, etc.
Some of these events are accompanied by GW memory \cite{favata2010memory}, which leaves permanent deformations of the spacetime after the passage of the GW burst signal. The pre-fit timing residual (PTA signal) induced by a GW memory event contains a ramp-like component, resulting in high PTA sensitivity \cite{madison2014merger,vanhaasteren2010memory}. For the sources and waveforms considered in this work, both the burst and monotonic memory are present and the latter develops over longer timescales.
\par In this paper, we provide a prescription, influenced by Refs.~\cite{susobhanan2023eccentric}, to efficiently search for burst-with-memory events associated with SMBH encounters in general relativistic hyperbolic orbits (we refer to such events as linear GW memory (LGWM) events in this paper).
It's worth emphasizing that linear memory essentially arises from certain non-oscillatory motion of a GW source in contrast with the non-linear memory that is associated with hereditary contributions to GW emission~\cite{lentati2013hyper, thorne1992gravitational}. The linear GW memory events can arise from binaries in hyperbolic orbits, the asymmetric ejection of masses or neutrinos, as inferred from detailed numerical studies~\cite{favata2010memory, richardson2022modeling}.
The PTA signals induced by such LGWM events were computed in Ref.~\cite{dandapat2023gravitational} employing a 3PN-accurate Keplerian-type parametric solution detailed in Ref.~\cite{cho2018hyperbolic} and the GW phasing approach of Ref.~\cite{damour2004phasing}. 
These efforts allowed us to describe general relativistic trajectories of compact objects in unbound orbits and compute the resulting temporally evolving GW polarisation states.
We recall that the PN approximation provides general relativistic corrections to the Newtonian prescription for hyperbolic orbital dynamics in terms of an expansion parameter $x = ( v/c)^2 \sim G\,M/(r \, c^2) \ll 1$, where  $v$, $M$, and $r$ are respectively the orbital velocity, total mass, and relative separation of the binary, and the 3PN order includes all contributions that are accurate up to $(v/c)^6$. 
The detailed investigation of Ref.~\cite{dandapat2023gravitational} showed that the 3PN approximation is appropriate for describing such events if their peak frequencies lie in the nanohertz range.  Further, such events involving $10^9\,M_{\odot} $ SMBHs should be detectable by an SKA-era PTA to cosmological distances. 
This work explores the PTA implications of an improved prescription to model such LGWM events.

In what follows, we describe in detail a computationally efficient version of the approach presented in Ref.~\cite{dandapat2023gravitational} for modeling LGWM events due to hyperbolic encounters of black holes, influenced by Ref.~\cite{susobhanan2023eccentric}. 
Thereafter, we delve into the data analysis implications of this improved prescription in Sec.~\ref{inj:sec} while employing simulated PTA datasets. 
Further, we searched for such LGWM events in using NANOGrav 12.5-year dataset using 44 pulsars.
This section also provides a comprehensive overview of the underlying Bayesian pipeline and explores the implications of utilizing different samplers in our searches.
Due to the absence of any LGWM events in the NANOGrav 12.5-year dataset, we provide an upper limit for the signal amplitude and probe its dependencies with the characteristic frequency, sky location, and various other parameters.  
A summary and possible future directions are listed in Sec.~\ref{sum:dis}. 

\section{PN-accurate modeling of hyperbolic encounters involving black holes} 
\label{sec:orb:eval}

We begin by providing a brief description of our approach to obtain temporally evolving quadrupolar order $h_{+,\times}$ for non-spinning BH binaries in 3PN-accurate hyperbolic orbits, influenced by Ref.~\cite{dandapat2023gravitational}.  
A computationally efficient way of obtaining the resulting pulsar timing residuals $R(t)$ is described in Sec.~\ref{sec:LGWM_signals}. 
Various facets of the resulting pipeline \texttt{GWhyp} are provided in Sec.~\ref{sec:gwhyp}.

~

\subsection{Temporally evolving \texorpdfstring{$h_{\times,+}(t)$}{hpx(t)} for hyperbolic BHs}
\label{sec:hyperbolic_hpx}

We begin by writing down the quadrupolar order GW polarization states $h^{\rm Q}_{+,\times}$ for non-spinning compact binaries in non-circular orbits \cite{gopakumar2002waveform}
\begin{widetext}
\begin{subequations}
\begin{align}
    h^{\rm Q}_{+}&=-\frac{G M \eta}{c^4 \, D_L} \bigg{\{} (1+\cos^2\iota) \left[\left(\frac{G M}{r}+r^2\dot{\phi}^2-\dot{r}^2\right)\cos 2 \phi + 2 r \dot{r} \dot{\phi} \sin 2 \phi\right]
    +\sin^2 \iota \,  \left(\frac{G M}{r}-r^2 \dot{\phi}^2-\dot{r}^2\right)\bigg{\}}\,, \\
    h^{\rm Q}_{\times}&=- \, \frac{G M \eta}{c^4 \, D_L} 2 \cos \iota \bigg{\{} \left(\frac{G M}{r}+r^2 \dot{\phi}^2-\dot{r}^2\right) \sin 2 \phi-2 r \dot{r}\dot{\phi} \cos 2 \phi \bigg{\}} \,,
\end{align}
\label{eq:hpx_rphi}
\end{subequations}
\end{widetext}
where 
$M$, $\eta$, and $D_L$ stand for the total mass, the symmetric mass ratio, and the luminosity distance to the GW source, respectively.
The angle $\iota$ represents the constant orbital inclination, while the dynamical variables $r$ and $\phi$ stand for the relative orbital separation and the angular coordinate in the orbital plane ($\dot{r}$ and $\dot\phi$ represent the time derivatives of these variables).
We now customize the above expressions for BH binaries in hyperbolic trajectories with the help of a Keplerian-type parametric solution for hyperbolic trajectories.
Influenced by Ref. \cite{damour1985pn}, we write 
\begin{subequations}
\label{eq:orbitalp}
\begin{align}
r &= a_r \left ( e_r\,\cosh u  -1 \right )\,,\\
n\,(t-t_0) & = e_t\sinh u -u \,,\\
(\phi-\phi_0) &= (1+k) \nu,
\end{align}
\end{subequations}
where $t_0$ is a certain fiducial time such that $\phi_0 = \phi(t_0)$. 
The orbital elements $a_r$, $e_r$, $e_t$, $k$, and $n$ are the semi-major axis, the radial eccentricity, the time eccentricity, the rate of periastron advance, and the mean motion, which are functions of the orbital energy and angular momentum, and $u$ and $\nu$ are the eccentric anomaly and the true anomaly \cite{cho2018hyperbolic}. 
The explicit expression for the true anomaly in terms of $u$ reads
\begin{equation}
    \nu = 2\arctan\left[\sqrt{\frac{e_\phi+1}{e_\phi-1}}\tanh \frac{u}{2}\right]\,,
    \label{eq:termnu}
\end{equation}
where $e_\phi$ is known as the angular eccentricity.

To obtain $h^{\rm Q}_{\times,+}$ for compact binaries in hyperbolic trajectories, we replace $r$, $\dot r$, and $\dot \phi$ appearing in Eqs.~\eqref{eq:hpx_rphi} with the help of Eqs.~\eqref{eq:orbitalp}, and this leads to
\begin{widetext}
\begin{subequations}
\label{qdh2}
\begin{align}
h^{\rm Q}_+ &=
-\mathcal{H} \, \bigg{[} (c_\iota^2+1)\bigg{(} \frac{2e_t^2-\chi^2+\chi-2}{(\chi-1)^2} \, \cos 2 \phi+\frac{2 \, \sqrt{e_t^2-1}}{(\chi-1)^2} \, \sin 2 \phi \bigg{)}  -s_\iota^2 \frac{\chi}{\chi-1} \bigg{]}\,,\\
h^{\rm Q}_\times &=
\mathcal{H} \, 2 c_\iota \, \bigg{[} \frac{2 \sqrt{e_t^2-1} \, \xi}{(\chi-1)^2} \, \cos 2 \phi-\frac{2e_t^2-\chi^2+\chi-2}{(\chi-1)^2} \, \sin 2 \phi \bigg{]}\,,
\end{align}
\label{eq:h_uphi}
\end{subequations}
\end{widetext}
where $\mathcal{H}=\frac{GM\eta}{D_L \, c^2}  \, x$ is the GW amplitude and $x=(GMn/c^3)^{2/3}$ is a dimensionless PN parameter. 
The additional notations used here include 
$c_\iota=\cos \iota$, $s_\iota=\sin \iota$, $\chi=e_t  \cosh u$, and $\xi=e_t  \sinh u$.
It should be noted that 
the Keplerian parametric solution (Eqs.~\ref{eq:orbitalp}) allows us to express  $ \dot \phi =  ( d \phi /d v) \times ( d v/du) \times ( d u/dl) \times ( dl/dt) $ and $\dot r = ( d r/du ) \times ( du/dl) \times ( dl/dt) $ in Eqs.~\eqref{eq:hpx_rphi}.
Further, we write the energy and angular momentum, required to specify the orbital elements, in terms of $n$ and $e_t$, while identifying orbital eccentricity with the time-eccentricity  $e_t$ that enters the PN-accurate Keplerian type parametric solution for hyperbolic orbits \cite{cho2018hyperbolic}.
The structure of Eqs.~\eqref{eq:h_uphi} allows us to pursue certain restricted PN-accurate prescription for modeling the temporal evolution of quadrupolar order GW polarisation states,
associated with non-spinning compact objects in PN-accurate hyperbolic orbits. 
In other words, we will employ a PN-accurate prescription for the time evolution of $u$ and $\phi$ variables that appear in Eqs.~\eqref{eq:h_uphi}. 
This PN-accurate evolution is implemented by employing the 
3PN-accurate Keplerian type parametric solution that extends Eqs.~(\ref{eq:orbitalp}-\ref{eq:termnu}), and detailed in Refs.~\cite{cho2018hyperbolic,dandapat2023gravitational}. The temporal evolution of eccentric anomaly $u$ is provided by the 3PN-accurate Kepler equation, which we write  explicitly as  \cite{dandapat2023gravitational}
\begin{widetext}
\begin{align}
l= n\, (t- t_0) &=(e_t \sinh u-u)-\frac{x^2 (12 \nu (-5+2 \eta )+e_t (-15+\eta ) \eta  \sin \nu)}{8 \sqrt{e_t^2-1}}+\frac{x^3}{6720 (e_t^2-1)^{3/2}}
\bigg{\{}e_t (67200\notag \\[2ex]
&\qquad-3 (-47956+105 e_t^2+1435 \pi ^2) \eta -105 (592+135 e_t^2) \eta ^2+35 (-8+65 e_t^2)
\eta ^3) \sin \nu\notag \\[2ex]
&\qquad+35 ( (8640-13184 \eta +123 \pi ^2 \eta +960 \eta ^2+96 e_t^2 (30-29 \eta +11 \eta ^2)) \, \nu+12
e_t^2 \eta  (116-49 \eta +3 \eta ^2) \notag \\[2ex]
&\qquad \times \sin 2 \nu+e_t^3 \, \eta  (23-73 \eta +13 \eta ^2) \sin \ 3 \nu)\bigg{\}} \,.
\label{eq:kepler_full}
\end{align}   
\end{widetext}
The explicit 3PN-accurate expression for $l$ is listed as 
to explain our efficient approach to tackle 
the PN-accurate Kepler equation.
Further, 
we employ the 3PN-accurate  expression for $\phi$, expressed in terms of $u$ and $\nu$ and adapted from Ref.~\cite{dandapat2023gravitational}
and we write  
symbolically as
 \begin{align}
    \phi=\phi_0+(1+k) \, \nu+\mathscr{F}_\phi(u) \, ,
    \label{eq:termphi}
\end{align}
where $k$  
is the advance of periapsis per orbit, and $\mathscr{F}_\phi(u)$ stands for the 2PN and 3PN periodic contributions to $\phi$
as detailed in Ref.~\cite{dandapat2023gravitational}.

It should be now obvious that we need to solve the above 3PN-accurate Kepler equation.
to obtain the PN-accurate temporal evolution for $h^{\rm Q}_{+,\times}$, given by Eqs.~\eqref{eq:h_uphi}.
It is customary to employ Mikkola's method to solve the classical Kepler equation,
namely $ l=e_t \, \sinh u -u$, as this method is the most accurate and efficient approach to tackle such a transcendental equation~\cite{mikkola1987cubic}.
We need additional steps to tackle our PN-accurate Kepler equation which can be
written symbolically as 
\begin{align}
    l=e_t \, \sinh u -u +\mathscr{F}_t(u) \, ,
\end{align}
where the explicit 2PN and 3PN contributions to the orbital function $\mathscr{F}_t(u)$ can easily be extracted from Eq.~\eqref{eq:kepler_full}.
In order to obtain 
the benefits of Mikkola's approach, 
re-write the PN-accurate Kepler equation in the form of the classical Kepler equation. 
For this purpose, we  introduce an 
`auxiliary eccentric anomaly' $\hat{u}$, inspired by Ref.~\cite{cho2022generalized}, which allows us to express the PN-accurate Kepler equation as
\begin{equation}
 l= \hat{u}- e_t\, \sinh \hat u\,.
\label{eq:kepler}
\end{equation}
The explicit PN-accurate expression for $u$ in terms of $\hat u$, relevant for the present effort, is given by Eq.~(27) of Ref.~\cite{dandapat2023gravitational}.
We can now employ Mikkola's method and obtain $\hat u$ as a function of $l$.
Thereafter, we obtain $u$ in terms of $l$ by invoking a PN-accurate expression that expresses $u$ in terms of $\hat u$, as noted earlier.
Thereafter, we can easily obtain $\phi (l)$ with the help of Eq.~\eqref{eq:termphi}. 
It should be noted that this procedure naturally allows us to obtain 
$h^{\rm Q}_{\times,+} (t)$ due to 3PN-accurate conservative orbital dynamics of 
two BHs in hyperbolic orbits in the restricted PN-accurate 
approach in 
a computationally efficient way.

In what follows, we briefly list how we incorporate the effects of GW emission that occurs at the 2.5PN order, influenced by Refs.~\cite{damour2004phasing,cho2018hyperbolic}.
Under the influence of GW emission, the unbound binary loses both the energy and the angular momentum which ensures that both $n$ and $e_t$ become time-dependent.
Further, the definition of the mean anomaly $l$ becomes 
\begin{align}
    l(t)=\int_{t_0}^t \, n(t^\prime) dt^\prime \, .
\end{align}
With these considerations, it is fairly straightforward to obtain the following  equations for $x$, $e_t$, and $l$~\cite{dandapat2023gravitational}:
\begin{subequations}
\label{gw_em}
\begin{align}
\frac{d x}{dt} &=\frac{16}{15} \, \frac{c^3 \, x^5 \, \eta}{G \, M \, \beta^6} 	 \bigg{\{} 35 \, (1-e_t^2)+(49 -9 e_t^2)\, \beta+32 \, \beta^2 \notag \\
&\qquad+ 6\, \beta^3 \bigg{\}}
\,,\\
\frac{d e_t}{dt} &= \frac{8}{15} \, \frac{(e_t^2-1) \, x^4 \, c^3\, \eta}{G \, M \, e_t \, \beta^6} \bigg{\{} 35 \,  (1-e_t^2)+(49-9 \, e_t^2) \, \beta \notag \\
&\qquad+17 \, \beta^2+3 \beta^3 \bigg{\}}
\,,\\  
\frac{d l}{dt} &= n \, ,
\end{align}
\end{subequations}
where $\beta=e_t \, \cosh u-1$. 


Following Ref.~\cite{dandapat2023gravitational}, we obtain temporally evolving $h^{\rm Q}_{\times,+}$  due to two non-spinning BHs in fully 3PN-accurate hyperbolic orbits with the help of the following steps.
First, we specify the total mass $M$ and the mass ratio $q$ along with an impact parameter $b$ and eccentricity parameter $e_t$ at an initial epoch $t_0$ where the orbital phase value is $\phi_0$.
Thereafter, we obtain the associated mean motion $n$ by numerically inverting the 3PN accurate expression that expresses $b$ as a function of  $n, e_t$ and $\eta$ as given in Eq.~(9) of Ref.~\cite{dandapat2023gravitational}.
We now tackle the 3PN-accurate Kepler equation with these inputs and obtain the $u_0$ associated with the initial $l$ which leads to $h^{\rm Q}_{\times,+}$ at the initial epoch $t_0$ for specific and constant $\iota$.
The temporal evolution now follows by evolving the differential equations for $x, e_t$, and $l$, given by Eqs.~\eqref{gw_em} and repeating the above-listed steps at the new epoch $t_0 + \delta t$.

In what follows, we provide an improved way to compute pulsar timing residuals associated with such temporally evolving $h^{\rm Q}_{\times,+}(t)$.

\subsection{Computationally Efficient Approach to Model  \texorpdfstring{$R(t)$}{R(t)} due to Burst With Linear Memory Events}
\label{sec:LGWM_signals}

A GW signal passing across the line of sight between the observer and a pulsar induces time-varying modulations in the observed times of arrival (TOAs) of the pulsar's pulses \cite{lee2011gravitational}.
For a transient event, we may write 
\begin{align}
    R(t_E)=\int_{t_0}^{t_E} \, h(t_E^\prime)  \, dt^\prime_E \, ,
    \label{Eq12l}
\end{align}
where $t_0$ is a fiducial epoch and both $t_E$ and $t_0$ are measured in the solar system barycenter (SSB) frame. 
It should be noted that we have ignored the pulsar contributions that typically appear in the $R(t)$ expression for individual continuous PTA sources \cite{vanhaasteren2010memory}. 
This is because the maximum duration of these burst events is expected to be around a decade or so, and therefore such events are not expected to induce observable pulsar term contributions since the pulsars are typically located at $\sim$kilo parsec distances from the Earth. 
In other words, there are essentially no pulsar contributions to timing residuals in general while dealing with our bursts with memory events.
In highly unusual scenarios, when the sky position of the GW source closely aligns with that of a pulsar in the array, the contributions of the pulsar terms to the GW-induced timing residual for that pulsar can be non-negligible. 
However, in these cases, the pulsar term contribution will exhibit high covariance with the timing model of the pulsar.
Consequently, the decision to ignore the pulsar terms should not significantly impact our results.


We proceed by defining the dimensionless gravitational wave strain $h(t)$ that appear in Eq.~\eqref{Eq12l}  as
\begin{align}
h&=\begin{bmatrix}F_{+} & F_{\times}\end{bmatrix}\begin{bmatrix}\cos2\psi & -\sin2\psi\\
\sin2\psi & \cos2\psi
\end{bmatrix}\begin{bmatrix}h_{+}\\
h_{\times}
\end{bmatrix}\,,
\label{eq:h_pta}
\end{align}
\newline
where the antenna pattern functions $F_{+,\times}$ depend on the sky positions of both the pulsar and the GW source and their explicit expressions are available in Ref.~\cite{lee2011gravitational}.
Additionally, $\psi$ represents the GW polarization angle, and $h_{+,\times}$ represents the quadrupolar order GW polarization states given by Eqs.~\eqref{qdh2}.
The resulting  $R(t)$ may be expressed in terms of two temporally evolving functions $s_{+,\times}(t)$ such that 
\begin{align}
    R(t_E) &= \begin{bmatrix}F_{+} & F_{\times}\end{bmatrix}\begin{bmatrix}\cos2\psi & -\sin2\psi\\
\sin2\psi & \cos2\psi
\end{bmatrix}\begin{bmatrix}s_{+}(t_E)\\
s_{\times}(t_E)
\end{bmatrix}\,.
\label{eq:R(t_E)_transient}
\end{align}
Here, we have defined
\begin{align}
\label{eq:st}
  s_{\times,+}(t_E) & =   \int_{t_0}^t \, h^{\rm Q}_{\times,+} (t')\, dt'\,,
\end{align}
and in what follows, we provide a computationally efficient approach for evaluating these integrals.

It may be noted that we had obtained $s_{\times+}(t_E)$ by employing explicit numerical integration of these definite integrals in Ref.~\cite{dandapat2023gravitational}.
This step turned out to be the most computationally expensive part of the resulting \texttt{GW\_hyp}~\footnote{\url{https://github.com/subhajitphy/GW_hyp}} package. 
This prompted us to explore the possibility of tackling these integrals analytically, influenced by Ref.~\cite{susobhanan2023eccentric}, by pursuing the following steps.
First, we change the integration variable from $t$ to $u$ by employing the Kepler equation, which leads to $ dt = dl/n =  du \left ( e_t \, \cosh u-1\right )/n $. 
Thereafter, we write $\phi=\nu+\omega'$ where $\nu$ is given by Eq.~\eqref{eq:termnu} while  $\omega'$ stands for all the PN corrections that appear in the 3PN-accurate expression for $\phi$ given by  Eq.~\eqref{eq:termphi}. 
We assume $\omega'$ to be a constant while performing the above integration, which is strictly true only for Newtonian hyperbolic orbits.
This leads to 
\begin{widetext}
\begin{subequations}
\label{eq:ana}
\begin{align}
s_+^A(t)&=-\mathcal{S} ((c^2_\iota+1)(-\mathcal{P} \sin 2 \omega^\prime+\mathcal{Q} \cos \, 2 \omega^\prime)-s_\iota^2 \mathcal{R})\\
s_\times^A(t)&=-\mathcal{S} \, 2 c_\iota(\mathcal{P} \cos \, 2\omega^\prime+\mathcal{Q}\sin \, 2\omega^\prime), 
\end{align}   
\end{subequations}
with 
 \begin{subequations}
\begin{align}
 \mathcal{P}&=\frac{\sqrt{e_t^2-1}(\cosh \, 2u-e_t \cosh \, u)}{e_t \cosh u -1} \\
 \mathcal{Q}&=\frac{((e_t^2-2)\cosh u+e_t)\sinh u}{e_t \cosh u -1} \\
 \mathcal{R}&=e_t \, \sinh u  \,,
\end{align}
 \end{subequations}
\end{widetext}
where $\mathcal{S}$ is the PTA signal amplitude given by $\mathcal{S}=\mathcal{H}/n$.
A few comments are in order here. 
Strictly speaking, the above expressions are valid only at the Newtonian order where $\omega'$ by definition is a constant.  
However, we employ 3PN-accurate expression for $\omega'$, easily extractable from Eq.~\eqref{eq:termphi}, while evaluating the expressions for the restricted PN-accurate description for $s_{\times+}(t)$.
Further, we have verified the following identity
\begin{align}
    h_{+,\times}=n \, \frac{\partial s_{+,\times}^A}{\partial l} \bigg{|}_{e_\phi \to e_t; n, e_t, \omega},
\end{align}
where the LHS is given by Eq.~\eqref{qdh2} and it provides an interesting consistency check for our approach to obtain a semi-analytic prescription for $R(t_E)$ associated with PN-accurate hyperbolic encounters of BHs.
It is important to note that this approach is influenced by Ref.~\cite{susobhanan2023eccentric} that provided an accurate and efficient way to compute $s_{+,\times} $ associated with BH binaries in precessing eccentric orbits.

\begin{figure*}
\centering  
\includegraphics[width=1\linewidth]{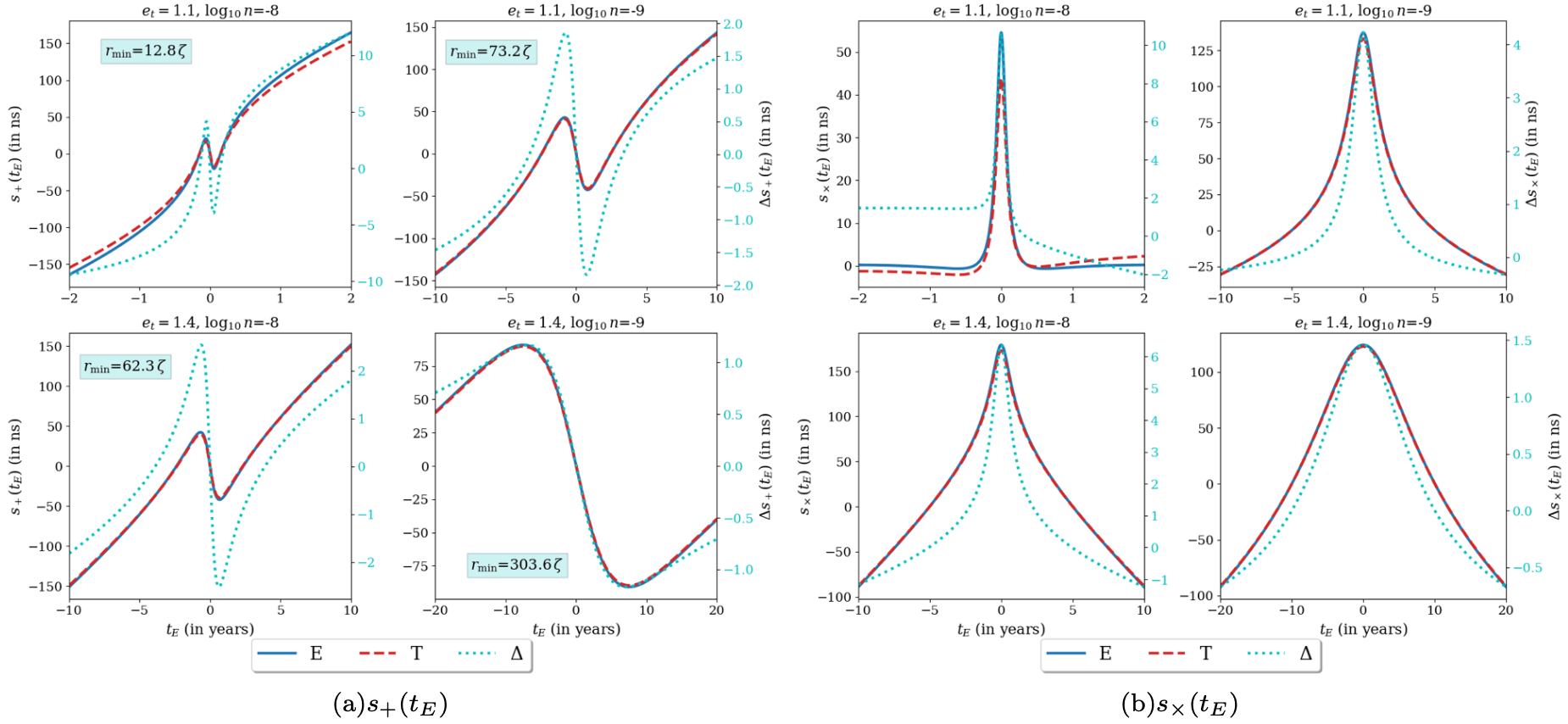} 
\caption{ 
Comparison between PTA signals that arise from our computationally efficient approach:~$s^E_{+,\times}(t_E)$  and the traditional approach:~$s^T_{+,\times}(t_E)$ 
for a system 
with $M=10^{10} M_\odot, \eta=1/4,  R^\prime=1.6 \text{ Gpc, and } \iota=\pi/3$ while 
varying  values of $\{e,\log_{10} n\}$ pairs 
Each panel displays three plots associated with the traditional, computationally efficient 
$s_{+}$ or $s_{\times}$ and their differences, labeled as $\Delta s$ with appropriate 
subscripts. The closest approach ($r_\text{min}$) between the BHs for each of these combinations is also displayed in terms of $\zeta=GM/c^2$. }
\label{diff_A_N}
\end{figure*}

We now compare the values of $s_{+,\times} $ arising from our earlier approach, detailed in Ref.~\cite{dandapat2023gravitational}, and the present approach. 
We refer to them as  $s^{\rm T}_{+,\times} $  and $s^{\rm E}_{+,\times} $ where `T' and `E' stand for the traditional and computationally efficient approaches. 
A detailed comparison of the PTA signals evaluated for different values of $n$ and $e_t$ using the two methods is shown in Fig.~\ref{diff_A_N}. 
The plots highlight that the distinction between the two approaches becomes apparent when the orbital separation is at the closest approach, for $r_\text{min} \sim 10 \, GM/c^2$. 
These deviations are expected because as $r_\text{min}$ decreases, the system becomes more relativistic, and our `E' approach is essentially an approximation of the `T' approach, which provides in principle a higher level of PN accuracy. 
However, we refrain from entering this $r_\text{min} \sim 10 \, GM/c^2$ regime, as explained in Section II.B of Ref.~\cite{dandapat2023gravitational} and also discussed in Ref.~\cite{cho2021instantaneous}, due to restrictions related to PN validity. 
Therefore, opting for the computationally efficient approach is justifiable, given that this approximation comfortably fits within the PTA parametric space and offers a substantial computational improvement ($\sim50$ times) over the traditional method.

In the next subsection, we provide details of the updated version of our \epr\cite{ellis_2020_4059815,johnson2023nanograv}-compatible \texttt{GWhyp} package and list its features. 

\subsection{Details of the \texttt{GWhyp} package and its features}
\label{sec:gwhyp}

\begin{figure}
\centering  
\includegraphics[width=1\linewidth]{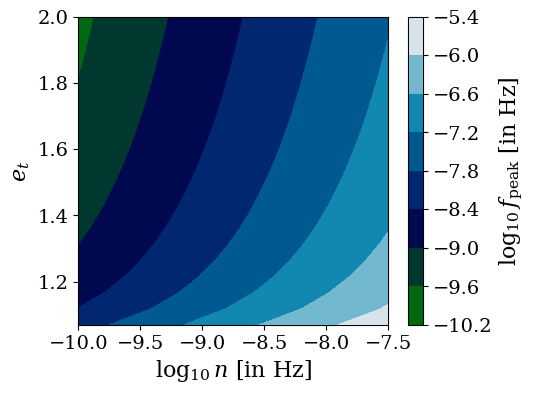}
\caption{A contour plot that relates the PTA-relevant peak GW frequencies ($f_\text{peak}$) of hyperbolic encounters with characteristic frequencies ($n$) and eccentricities ($e_t$).
It turns out that $n$ is a good proxy for $f_\text{peak}$ and it very weakly depends on $\eta$ and this plot is for equal mass BH encounters.}
\label{fp:n}
\end{figure}

We now present a detailed description of our \texttt{GWhyp} \footnote{\url{https://github.com/subhajitphy/GWhyp}} package which is the improved and updated version of our \texttt{GW\_hyp}~\cite{dandapat2023gravitational} Python package.  
The updated package implements the computationally efficient way of tackling the indefinite integrals present in the $s_{\times,+}(t_E)$ expression, given by Eqs.~\eqref{eq:st}.
Further, we employ the mean motion $n$ rather than the impact parameter $b$ to characterize the hyperbolic orbit, and this choice is based on two practical considerations. 
The first reason is that $n$ provides an excellent proxy to the peak frequency ($f_\text{peak}$) of the GW spectrum associated with hyperbolic encounters.
This conclusion arises from a detailed analysis of the Fourier domain expression for GW luminosity, $F_{\text Q}(f)$, given by Eq.~(3.27) of Ref.~\cite{bini2021frequency}.
It turns out that the $f_\text{peak}$ values, obtained by maximizing $F_{\text Q}(f)$, are close to $n$ values for a substantial part of the $\{n,e_t\}$ parameter space, relevant for our LGWM  sources.
It may be noted that we extracted $n$ values from a 3PN-accurate expression for $b$ in terms of $n$ and $e_t$, given by Eq.(9) of Ref.~\cite{dandapat2023gravitational}.
In Fig.~\ref{fp:n}, we display the way $f_\text{peak}$ relates to $n$ with the help of contour plots.
Interestingly, both $f_\text{peak}$ and $n$ exhibit a dependency on total mass $M$ via the scaling $c^3/GM$, resulting in an absence of $M$-dependence between them, for the relevant $e_t$ ranges. 
However, a rather weak dependence on $\eta$ is observed through the relationship that connects $b$, $n$, and $e_t$, given by in Eq.(9) of Ref.~\cite{dandapat2023gravitational}.
Additionally, $n$ is a more useful parameter in our opinion as it depends only on the conserved energy while $b$, similar to $e_t$, depends on both energy and angular momentum as evident from  PN-accurate expressions for these orbital elements available in Ref.~\cite{cho2018hyperbolic}.

Secondly, the usefulness of $n$ became more evident during our detailed parameter estimation studies, as described in Sec.~\ref{inj:sec}. 
It turned out to be very challenging to recover both $b$ and $e_t$ simultaneously in our injection studies; such parameter estimation runs only resulted in a constraint on  
one of these parameters rather than a full recovery.
In contrast, we regularly achieve simultaneous recovery of both $n$ and $e_t$ parameters while simultaneously searching over them in our injection studies. 
The above two considerations prompted us to employ $n$ and $e_t$ to characterize our PN-accurate hyperbolic orbits while providing ready-to-use routines to model $R(t)$ from such LGWM events. 

Our improved and updated \texttt{GWhyp} software package adheres to the following steps for calculating $R(t_E)$ associated with hyperbolic BH binaries with the total mass $M$ and mass ratio $q$ when provided with a collection of TOAs.
\begin{itemize}
    \item The function called \texttt{cal\_sp\_sx\_A} receives inputs such as PTA signal amplitude ($\mathcal{S}$), GW source coordinates, eccentricity ($e_t$), the above-described     characteristic frequency ($n$), 
    the inclination angle ($\iota$), and the fiducial time ($t_0$), along with the TOA values and pulsar coordinates.
    \item Thereafter, we solve the Kepler equation by 
    invoking Mikkola's method (Eqs.~\ref{eq:kepler}) 
    to obtain the eccentric anomaly $u$ for each value of $l$, $n$, and $e_t$.
    \item Subsequently, we calculate the 3PN-accurate $\omega^\prime$ using the available  $u$, $n$, $e_t$, and $q$ values.
    \item We pass the values of $\omega^\prime$, $u$, and $e_t$ into Eq.~\eqref{eq:ana} to obtain  both $s_+^E(t)$ and $s_\times^E(t)$ values.
    This leads 
    to 
    the desired timing residual $R(t_E)$ via Eq.~\eqref{eq:R(t_E)_transient}.
    \item Finally, we obtain PTA signal at TOAs via interpolation of dense $R(t)$ samples, utilizing the \texttt{scipy.interpolate.CubicSpline} class.
\end{itemize}

Furthermore, we have introduced a high-level function, labeled as `\texttt{GWhyp.hyp\_pta\_res\_A}', which generates PTA signals using an \texttt{ENTERPRISE} `\texttt{Pulsar}' object and a set of source parameters. 
This function can be easily utilized to generate an \texttt{ENTERPRISE} `\texttt{Signal}' object to search for GWs arising from hyperbolic encounters of SMBHs in the various PTA datasets.

\section{Probing PTA implications of our \texttt{GWhyp} package}
\label{inj:sec}

We now explain how we can employ our $R(t)$ prescription to search for GWs from hyperbolic encounters in PTA datasets and discuss its implications.
This effort is influenced by a very recent effort that pursued a Bayesian search for a possible eccentric SMBHB in an active galaxy 3C 66B using 44 pulsars in the NANOGrav 12.5-year dataset, detailed in  Ref.~\cite{Lokie2023nanograv}.
In what follows, we summarise how we adapt the above effort for our ready-to-use $R(t)$ prescription for individual hyperbolic events. 
Note that comprehensive discussions on Bayesian inference using PTA datasets can be found in, e.g., Refs.~\cite{arzoumanian2016nanograv,taylor2021nanohertz}.
We employ both simulated and actual PTA datasets in our investigations, and we begin by briefly discussing these datasets.


\subsection{Our Datasets}

The present study employs the NANOGrav 12.5-year (NG12.5) narrowband dataset \cite{alam2021ng12.5} to search for GWs originating from LGWM events.
This dataset encompasses the times of arrival (TOAs) and timing models for 47 pulsars, gathered between 2004 and 2017 using the Green Bank Telescope and the Arecibo Telescope. 
We excluded PSRs J1946+3417 and J2322+2057 from our analysis due to their limited observation duration (less than 3 years), while PSR J1713+0747 was omitted because it exhibited two chromatic timing events within the data span \cite{lam2018second}. 
The earliest and latest TOAs in the dataset were recorded on MJDs 53216 (July 30, 2004) and 57933 (June 29, 2017), respectively, resulting in an approximate data span of~$\sim$12.92 years. 
To account for dispersion measure (DM) variations within the TOAs, DMX parameters were applied to provide a piece wise constant model, and each TOA was transformed to the Solar System Barycenter (SSB) frame using the DE438 solar system ephemeris.

The simulated datasets used in our simulation studies are based on the NG12.5 dataset, containing fake TOAs corresponding to the same observing epochs and observing system configurations for the same pulsars, but with different noise properties and an injected LGWM signal.

The data analysis procedures applied to these datasets are explained in the forthcoming subsections.

\subsection{Modeling Noise and PTA Likelihood}

For PTA GW observatories, the main observables are the times of arrival (TOAs) of pulsar pulses, from which the timing residual ($\delta t$) can be derived with the help of a timing model \cite{hobbs2006tempo2}.
The timing residuals quantify the differences between the observed TOAs and those predicted by an appropriate deterministic pulsar timing model at each observing epoch.
The timing residuals typically consist of three main components: white noise, red noise, and small timing model deviations. 
A linearized model for the timing residuals can be written as
\begin{align}
    \vec{\delta t}=\vec{n}+\mathbf{F} \vec{a}+\mathbf{M} \vec{\epsilon} +\mathcal{R}.
\end{align}
where 
$\vec{n}$ denotes the time-uncorrelated (white) noise, $\mathbf{F}$ represents the Fourier basis matrix of time-correlated (red) noise with coefficients $\vec{a}$, and $\textbf{M}$ represents the pulsar timing design matrix containing partial derivatives of the timing residuals with respect to the timing model parameters.
$\vec{\epsilon}$ is a vector indicating minor deviations from the optimal timing model parameters and $\mathcal{R}$ stands for a GW-induced signal
(in the case of an LGWM event, this is given by Eq.~\ref{eq:R(t_E)_transient}).
We now briefly delve into each component of the aforementioned expression.

\paragraph{White Noise:}
For each TOA, the white noise is assumed to be a normally distributed random variable with zero mean, given by the expression:
\begin{align}
    \langle n_{i,\mu} n_{j,\nu} \rangle=E^2_\mu \sigma_i^2 \delta_{ij}\delta_{\mu \nu}+Q_\mu^2\delta_{ij}\delta_{\mu \nu}.
\end{align}
Here, $\sigma_i$ represents the TOA uncertainty in the $i$th observations, $E_\mu$ is the \textit{error factor} (EFAC) for the receiver-backend system $\mu$, and $Q_\mu$ is the \textit{error added in quadrature} (EQUAD). 
Additionally, ECORR parameters are used to represent the correlations among narrowband TOAs originating from the same observation~\citep{agazie2023nanogravd}.

\paragraph{Red Noise:}
The (achromatic) red noise refers to the slow stochastic wandering of TOAs usually attributed to the rotational irregularities of the pulsar, the GWB, etc.
It is typically modeled as a Fourier series with frequencies $j/T_{\text{span}}$, where the index $j$ represents the harmonic number, and $T_{\text{span}}$ is the observation time span. 
Red noise is a low-frequency phenomenon, so the summation over $j$ can be truncated to a reasonable cutoff $N_f$.
The Fourier basis matrix $\mathbf{F}$ contains a set of $2N_f$ sine-cosine pairs evaluated at different observation epochs corresponding to $N$ TOAs. 

The Fourier coefficients $\vec{a}$ are assumed to follow a normal distribution, having a mean of zero and a covariance matrix $\langle \vec{a} \, \vec{a}^T \, \rangle=\mathbf{\Phi}$. 
The covariance matrix $\mathbf{\Phi}$ incorporates all potential sources of low-frequency achromatic noise.
For the present analysis, we consider two possible sources: pulsar intrinsic red noise (IRN) and the GWB. 
This allows us to write $\mathbf{\Phi}$ as 
\begin{align}
    [\Phi]_{(ak)(bj)}=\Gamma_{ab}\rho_{k}^2\delta_{kj}+\kappa_{ak}^2\delta_{kj}\delta_{ab}\,.
\end{align}
Here, the first term represents the GWB and the second term represents the IRN.
The indices $a$ and $b$ represent the two pulsars, $k$ and $j$ denote the frequency harmonics, and $\Gamma_{ab}$ stands for the GWB overlap reduction function for the pulsar pair $(a, b)$. 
The weights $\rho_{k}^2$ are related to the GWB power spectral density $S(f)$ by the relation $\rho_{k}^2=S(k/T_{\text{span}})\,T_{\text{span}}^{-1}$. 
A similar relation also applies to $\kappa_{ak}^2$ and the power spectral density of the IRN.
It is customary to model these quantities using the following expressions:
\begin{subequations}
\begin{align}
    \rho^2(f)&=\frac{A_\text{GWB}^2}{12 \pi^2 T_{\text{span}}} \left( \frac{f}{1 \, \text{year}^{-1}} \right)^{-\gamma_\text{GWB}} \, \text{year}^2, \\
    \kappa_a^2(f)&=\frac{A_a^2}{12 \pi^2 T_{\text{span}}} \left( \frac{f}{1 \, \text{year}^{-1}} \right)^{-\gamma_a} \, \text{year}^2.
\end{align}
\end{subequations}

In our analysis, we model the IRN of a pulsar using a Fourier sum comprising 30 equally spaced frequency bins spanning from $1/T_\text{span}$ to $30/T_\text{span}$. 
Furthermore, we will employ a common uncorrelated red noise (CURN) model as opposed to a GWB model due to its computational efficiency (this corresponds to setting $\Gamma_{ab}\rightarrow\delta_{ab}$ instead of the Hellings-Downs overlap reduction function). 
This choice is supported by the findings of Ref.~\cite{arzoumanian2020nanograv} where they identified a CURN process in the NG12.5 dataset without any definite detection of spatial correlations. 
To align with the NG12.5 GWB analysis, we represent the CURN using a Fourier sum comprising five evenly spaced frequency bins, spanning from $1/T_\text{span}$ to $5/T_\text{span}$.

\paragraph{Small timing model deviations:}
The term $\mathbf{M}\vec{\epsilon}$ takes into consideration the potential minor deviations from the initial best-fit values of the $m$ timing-ephemeris parameters due to the introduction of the additional signal and noise components. 
The design matrix $\mathbf{M}$, is an $N \times m$ matrix that holds the partial derivatives of the $N$ TOAs with respect to each timing model parameter, evaluated at the initial best-fit value. 
$\vec{\epsilon}$ is a vector comprising of linear offsets from these initial best-fit parameters associated with the deterministic pulsar timing model.

~\\

In PTA analyses, the focus lies primarily on the statistical properties of the noise rather than its individual realizations, and we can marginalize over the values of $\vec{a}$ and $\vec{\epsilon}$. 
This leads to a marginalized likelihood, which solely depends on the hyper-parameters $(A_\text{GWB}, \gamma_\text{GWB}, . . .)$, representing the statistical characteristics of the noise~\citep{lentati2013hyper, van2013understanding}:
\begin{align}
    p(\vec{\delta t}|\eta)=\frac{\exp(-\frac{1}{2} \, \vec{\delta t}^T \, \mathbf{C}^{-1} \vec{\delta t})}{\sqrt{\det{(2 \pi \mathbf{C})}}},
\end{align}
where $\mathbf{C}=\mathbf{N}+\mathbf{T} \mathbf{B} \mathbf{T}^T$. 
Here, $\mathbf{N}$ refers to the covariance matrix associated with the white noise, $\mathbf{B}=\text{diag}(\infty, \mathbf{\Phi})$, and $\mathbf{T}=[\mathbf{M},\mathbf{F}]$.

\begin{table*}[t]

\begin{center}
\renewcommand{\arraystretch}{1.3}
\setlength{\tabcolsep}{6.5pt}
\small
\begin{tabular}{llll}

\hline \hline
\toprule
\multicolumn{1}{c}{\bf{Parameter}}  & \multicolumn{1}{c}{\bf{Description}} & \multicolumn{1}{c}{\bf{Prior}} & \multicolumn{1}{c}{\bf{Comments}} \\
\midrule

\multicolumn{4}{c}{\bf{Intrinsic Pulsar Red Noise (IRN)}} \\[1pt]
$\log_{10} A_{\rm red}$ & Red noise power-law amplitude & Uniform $[-20, -11]$ & One parameter per pulsar  \\
$\gamma_{\rm red}$ & Red noise power-law spectral index & Uniform $[0, 7]$ & One parameter per pulsar \\
\midrule

\multicolumn{4}{c}{\bf{Common Uncorrelated Red Noise (CURN)}} \\[1pt]
$\log_{10} A_{\mathrm{CURN}}$ & Common process strain amplitude & Uniform $[-20, -11]$ & One parameter for PTA \\
$\gamma_{\mathrm{CURN}}$ & Common process power-law spectral index & Uniform $[0, 7]$& One parameter for PTA \\

\multicolumn{4}{c}{\bf{\ghn: Shape Parameters}} \\[1pt]
$\log_{10} n$ [Hz] & \ghn characteristic frequency & $\text{Uniform}[-10, -6.5] \times \mathcal{V}(M, n, \epsilon, q)$ & One parameter for PTA \\
$t_\text{ref}$ [MJD] & Fiducial time of the LGWM event & $\text{Uniform}[53216,57933]$ & One parameter for PTA \\
$e_t$ & Orbital eccentricity  & $\text{Uniform}[1.08,2] \times \mathcal{V}(M, n, \epsilon, q)$ & One parameter for PTA \\
$\log_{10} M$ $[M_\odot]$ & Total mass of the system  & $\text{Uniform}[7, 10] \times \mathcal{V}(M, n, \epsilon, q)$  & One parameter for PTA \\
$q$ & Mass ratio of the binary system & $\text{Uniform}[0.1,1]\times \mathcal{V}(M, n, \epsilon, q)$  & One parameter for PTA \\
\multicolumn{4}{c}{\bf{\ghn: Projection Parameters}} \\[1pt]
$\log_{10} \, \mathcal{S}$ [s] & \ghn signal amplitude & $\text{Uniform}[-10, -5] / \text{LinearExp}[-10, -5]$ & One parameter for PTA \\
$\cos \iota$  & Cosine of the orbital inclination  & Uniform $[0, 1]$ & One parameter for PTA \\
$\alpha_\text{GW}$ (rad) & Right ascension of the GW source & Uniform $[0, 2\pi]$ & One parameter for PTA\\
$\delta_\text{GW}$ (rad) & Declination of the GW source & Uniform $[0, 2\pi]$ & One parameter for PTA\\
$\psi_\text{GW}$ (rad) & GW polarization angle  & Uniform $[0, \pi]$ & One parameter for PTA\\
\midrule

\bottomrule
\hline \hline

\end{tabular}

\caption{
Details of the 
prior distributions for various parameters that are relevant for various 
searches.
\label{param:tab}}
\end{center}

\end{table*}

\subsection{Details of the Bayesian Pipeline and the Employed Priors}

We employ the \ghn package to compute the PTA signal induced by linear GW memory events and utilize the \epr\cite{ellis_2020_4059815,johnson2023nanograv} package to evaluate the relevant likelihood function and the prior distribution (we will describe our priors in this section).
\subsubsection{Model comparison}
We utilized the Savage–Dickey formula to calculate Bayes factors indicating the existence of a gravitational wave signal~\cite{dickey1971weighted},
\begin{align}
    \mathcal{B}_{10}=\frac{\text{evidence}[\mathcal{H}_1]}{\text{evidence}[\mathcal{H}_0]}=\frac{p(\mathcal{S}=0|\mathcal{H}_1)}{p(\mathcal{S}=0|\mathcal{D},\mathcal{H}_1)},
\end{align}
where $\mathcal{H}_1=$Timing Model$+$WN$+$IRN$+$CURN$+$LGWM and $\mathcal{H}_0=$Timing Model$+$WN$+$IRN$+$CURN, $p(\mathcal{S}=0|\mathcal{H}_1)$ is the prior volume at $\mathcal{S}=0$, and $p(\mathcal{S}=0|\mathcal{D},\mathcal{H}_1)$ is the posterior volume at  $\mathcal{S}=0$. 
We applied the Savage–Dickey formula due to the nested nature of models $\mathcal{H}_1$ and $\mathcal{H}_0$, with $\mathcal{H}_0$ being a special case of $\mathcal{H}_1$ when $\mathcal{S}=0$. 
Our approximation for $p(\mathcal{S}=0|\mathcal{D},\mathcal{H}_1)$ involved estimating it as the ratio of quasi-independent samples within the lowest-amplitude bin of  $\mathcal{S}$ histogram. The uncertainty in this estimation is determined by ${\mathcal{B}_{10}}/{\sqrt{N_0}}$, with $N_0$ as the number of samples in the lowest amplitude bin.

We now direct our attention to specifying priors and conducting analyses on both simulated and actual datasets. 

\subsubsection{Specifying the prior distribution}
For present efforts, we provide specific prior distributions for the LGWM, IRN, and CURN parameters as listed in Table~\ref{param:tab}.
It should be noted that we employ two different types of priors for the LGWM signal amplitude ($\mathcal{S}$), namely a uniform prior on $\log_{10}\mathcal{S}$ for detection analysis and a uniform prior on $\mathcal{S}$ itself for upper limit analysis (`LinearExp' in Table~\ref{param:tab}). 
Although $q\in(0,1]$ and $e_t \in (1,\infty)$ by definition, we set small positive lower bounds for $q$ and $e_t-1$ to prevent numerical errors such as division by zero.  
Further,  we fix the upper limit for $\psi$ at $\pi$ as it enters the PTA signal through $\sin 2 \psi$ and $\cos 2 \psi$. It's worth mentioning that during our real search, we fixed the White Noise parameters from the previous NANOGrav 12.5-year noise analysis results~\cite{arzoumanian2020nanograv}.

\begin{figure*}
\centering  
\includegraphics[width=0.6\linewidth]{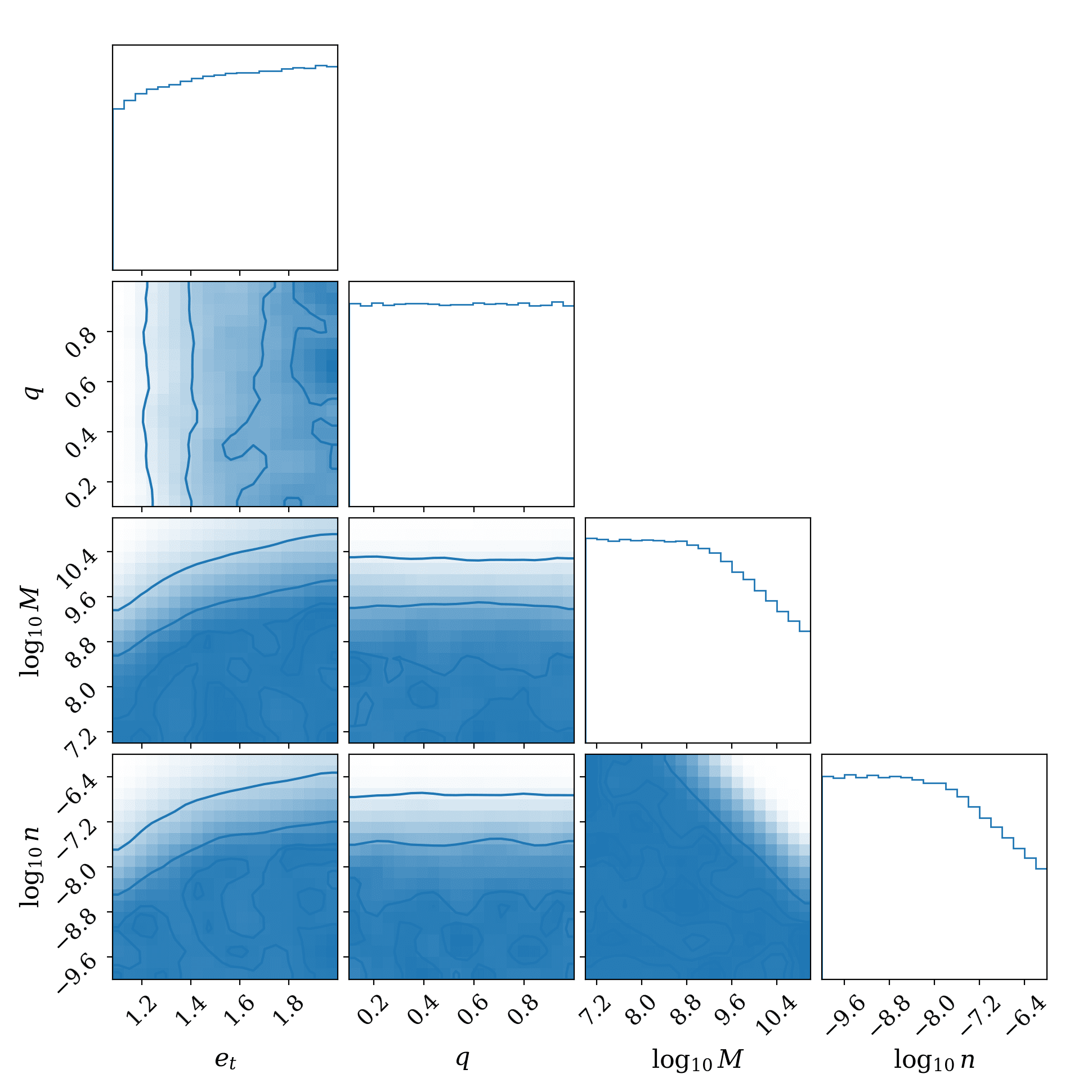}
\caption{The influence of imposing the validation criterion $r_{\rm min}  > 10 \, G\,M/c^2 $ on the joint prior distribution of $M$, $e_t$, $n$, and $q$ as $r_{\rm min}$ depends on these parameters in our PN-accurate approach. 
This corner plot displays how the uniform prior distribution on these parameters gets modified when we impose certain validation function $\mathcal{V}(\log_{10}M, n, e_t, q)$ to implement the $r_{\rm min}  > 10 \, G\,M/c^2 $ restriction. The samples from the prior are drawn through rejection sampling.}

\label{valid_plot}
\end{figure*}

Influenced by Ref.~\citep{Lokie2023nanograv}, we assign zero prior probability to parameter combinations that do not meet the following criterion: the orbital separation at the closest approach between the BHs should be $r_\text{min} \gtrsim 10 \, GM/c^2$ throughout the evolution of the binary within the data span.
This is to ensure the validity of PN approximation as detailed in \cite{dandapat2023gravitational,cho2021instantaneous}.
We employ a validation function $\mathcal{V}(M, n, e_t, q)$ which is set to $1$ if the values of $(M, n, e_t, q)$ meet the above criterion, and 0 otherwise. 
This function modifies the uniform prior distributions for $M$, $n$, $e_t$, and $q$ as given in Table~\ref{param:tab}, ensuring a zero prior when our PTA signal description is inadequate from the PN perspective.
This ultimately alters the joint prior distribution of $M$, $n$, $e_t$, and $q$, as shown in Fig~\ref{valid_plot}.


\subsubsection{ Exploring the implications of various samplers}

When conducting injection studies on simulated datasets, we employ the \nau\cite{lange2023nautilus} sampler, while for real searches on the NANOGrav 12.5-year dataset, we utilize the \ptm~\cite{ellis2017jellis18,johnson2023nanograv}.
A few comments are appropriate for introducing the \nau package. 
It may be noted that we employ this open-source \texttt{Python}-compatible package for Bayesian posterior and evidence estimation.
Interestingly, the \nau sampler employs a novel approach to enhance the efficiency of the Importance Nested Sampling (INS)~\cite{feroz2013importance} technique for Bayesian posterior and evidence estimation through deep learning. 
In lower dimensions ($\lesssim 50$), \nau consistently demonstrates significantly higher sampling efficiency compared to most of other samplers, including the traditional \ptm~commonly used in PTA studies, often exceeding it by more than an order of magnitude. 
Additionally, \nau produces highly accurate results with fewer likelihood evaluations compared to \ptm. 
In Fig.~\ref{inj_NG12.5_44psr_compare_NP} (to be discussed in detail in Section~\ref{simu:44psr}), we illustrate the parameter estimation consistency between the \ptm~and \nau sampler, revealing similar posterior recovery results using the Raveri-Doux tension metric \cite{raveri2021non}, which turned out to be  $<0.01 \, \sigma$.
Notably, the computational time for obtaining these results was only about 50 minutes for \nau, whereas it took around 2.5 days for the \ptm.
\par

However, when we are searching for linear GW memory signals on the real NANOGrav 12.5-year dataset, it is essential to consider the intrinsic red noise parameters (IRN) for each pulsar, leading to an increased dimensionality of 2 per pulsar. 
For instance, in the real NANOGrav 12.5-year dataset comprising 44 pulsars, the total search dimensionality becomes $44\times2$ (IRN) + $10$ (LGWM) + 2 (CURN) = 100. 
In such scenarios, one of the reliable samplers remains \ptm, as the Nested Sampling algorithm breaks down in this high dimension.
Nevertheless, when conducting injection studies in lower dimensions ($\lesssim 50$) to validate our search pipeline, \nau remains one of the most effective samplers.

During the actual NG12.5-year search using the \ptm, we assign weights of 25, 20, and 15 to the adaptive Metropolis (AM), single-component adaptive Metropolis (SCAM), and differential evolution (DE) proposal distributions, respectively. We employ various proposals provided within the \texttt{enterprise\_extensions}~\cite{TaylorBaker+2021, johnson2023nanograv} package. These include sampling from single-parameter priors (\texttt{JumpProposal.d\\raw\_from\_prior}, with a weight of 10, and \texttt{JumpProposal\\.draw\_from\_par\_prior}, with a weight of 2 for the CURN parameter), as well as sampling from empirical distributions (\texttt{JumpProposal.draw\_from\_empirical\_distr}), with a weight of 30 assigned to IRN parameters. In addition, we use parallel tempering with four geometrically spaced temperatures.

\subsection{ Injection Studies using Simulated datasets} 
\label{simu:44psr}
We now apply the above two approaches to simulated datasets containing our LGWM events to evaluate the consistency and effectiveness of our methods. 
We note that it is not possible to estimate the projection parameters $(\log_{10} \mathcal{S},\cos\iota,\psi_{\text{GW}})$ and the source coordinates $(\alpha_{\text{GW}},\delta_{\text{GW}})$ simultaneously and independently using only one pulsar \citep{susobhanan2023eccentric}.
Further, due to the low number of GW cycles, extracting all LGWM parameters as listed in Table~\ref{param:tab} simultaneously should be difficult while only employing one pulsar (to be thoroughly discussed in Appendix~\ref{app:gwecc:gwhyp}).
This naturally forces us to employ multiple pulsars and we employ simulated datasets based on 44 pulsars from the NG12.5 GWB search~\cite{arzoumanian2020nanograv}.
We generate the relevant TOAs using the \texttt{libstempo} package ~\cite{vallisneri2020libstempo}, which is a \texttt{Python} wrapper for the widely used \texttt{TEMPO2} package~\cite{hobbs2006tempo2}.

We begin by creating ideal noise-free realizations of pulsar TOAs by subtracting timing residuals from the measured TOAs obtained from the NG12.5 narrow-band dataset. 
We then inject White Noise (WN) with the same measurement uncertainties as the NG12.5 narrow-band dataset, with EFAC, EQUAD, and ECORR fixed at their maximum likelihood values derived from single-pulsar noise analysis, following standard PTA practices. 
Additionally, we inject a realization of the Common Uncorrelated Red Noise (CURN) with $A_\text{CURN}=2.4\times10^{-15}$ (the GWB amplitude estimated from the NANOGrav 15-year dataset~\cite{agazie2023nanograv}) and $\gamma_\text{CURN}=13/3$, in line with the expected characteristics of a GWB originating from  SMBHBs inspiraling along circular orbits due to the emission of nHz GWs
~\cite{phinney2001practical}.

\begin{figure*}[t!]
    \centering
    \includegraphics[width=0.9\textwidth]{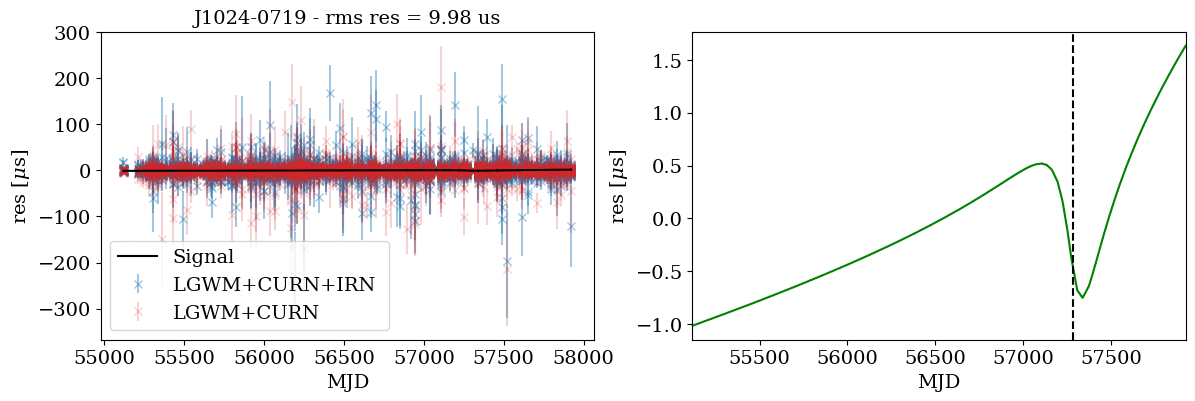} 
  \includegraphics[width=0.9\textwidth]{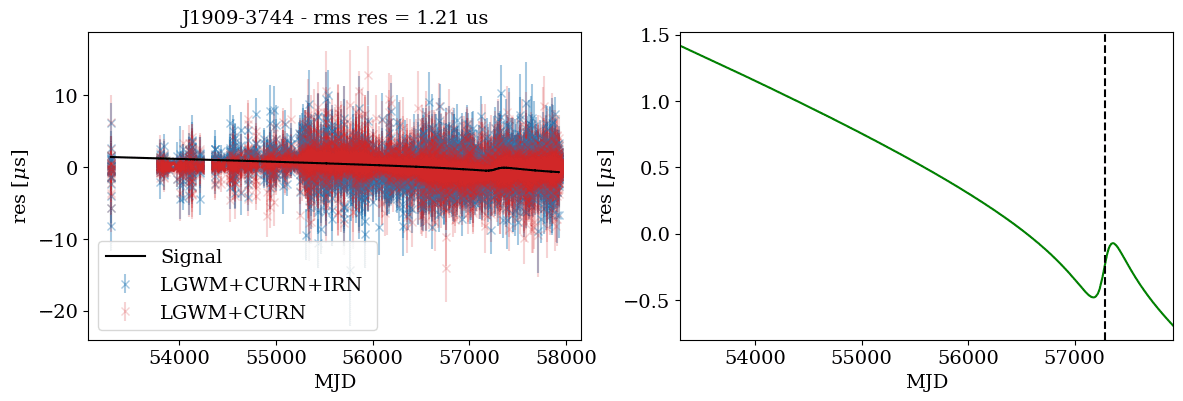}   
      \caption{ Plots for injected timing residuals containing LGWM signal with strength of 1$\mu s$. We consider  noisy (J1024-0719) and less noisy (J1909-3744) pulsars for two scenarios: LGWM+CURN and LGWM+CURN+IRN. In these plots, the left side shows injected residuals with the LGWM signal, while the right side shows the actual LGWM signal.
      Our bare signal, characterised by $\log_{10} \mathcal{S}$,
      is more evident in the less noisy pulsar. The vertical line in the right plots refers to the burst epoch ($t_\text{ref})$.
      }
       \label{fig:snrs}
\end{figure*}

We are now in a position to pursue injection studies of our LGWM signals, characterized by specific values of the `shape' and `projection' parameters listed in Table.~\ref{param:tab}, using the above-described simulated TOAs.
Our studies deal with two distinct scenarios.
In the first set of runs, we do not inject Intrinsic Pulsar Red Noise (IRN) realizations into our simulated data mainly due to simplicity and computational efficiency considerations. 
In the second set of runs, we include IRN realizations injected at their maximum likelihood values obtained from the NG12.5 GWB search.
In what follows, we refer to these two categories as LGWM+CURN and LGWM+CURN+IRN searches. It should be noted that we employ maximum-likelihood estimation to fit the initial timing model to the signal-injected simulated TOAs in both of these cases, saving the resulting post-fit timing model and TOAs as \texttt{par} and \texttt{tim} files. We gather from these studies that the strength of these LGWM signals is essentially provided
by their amplitude 
conveniently characterized by $\log_{10} \mathcal{S}$,  as listed in Table.~\ref{param:tab}. 
 In Fig.~\ref{fig:snrs}, we plot our signals are on the top of injected timing residuals for  specific pulsars.
We inject a LGWM signal with an amplitude of 1$\mu s$ ($\log_{10} \mathcal{S}=-6$) for the LGWM+CURN and LGWM+CURN+IRN scenarios, considering both noisy (J1024-0719) and less noisy (J1909-3744) pulsars. These visualizations demonstrate that our signal becomes more evident for less noisy pulsars, indicating that the above amplitude serves as a proxy for the underlying signal-to-noise ratio.

\par
\begin{figure*}
\centering  
\includegraphics[width=1\linewidth]{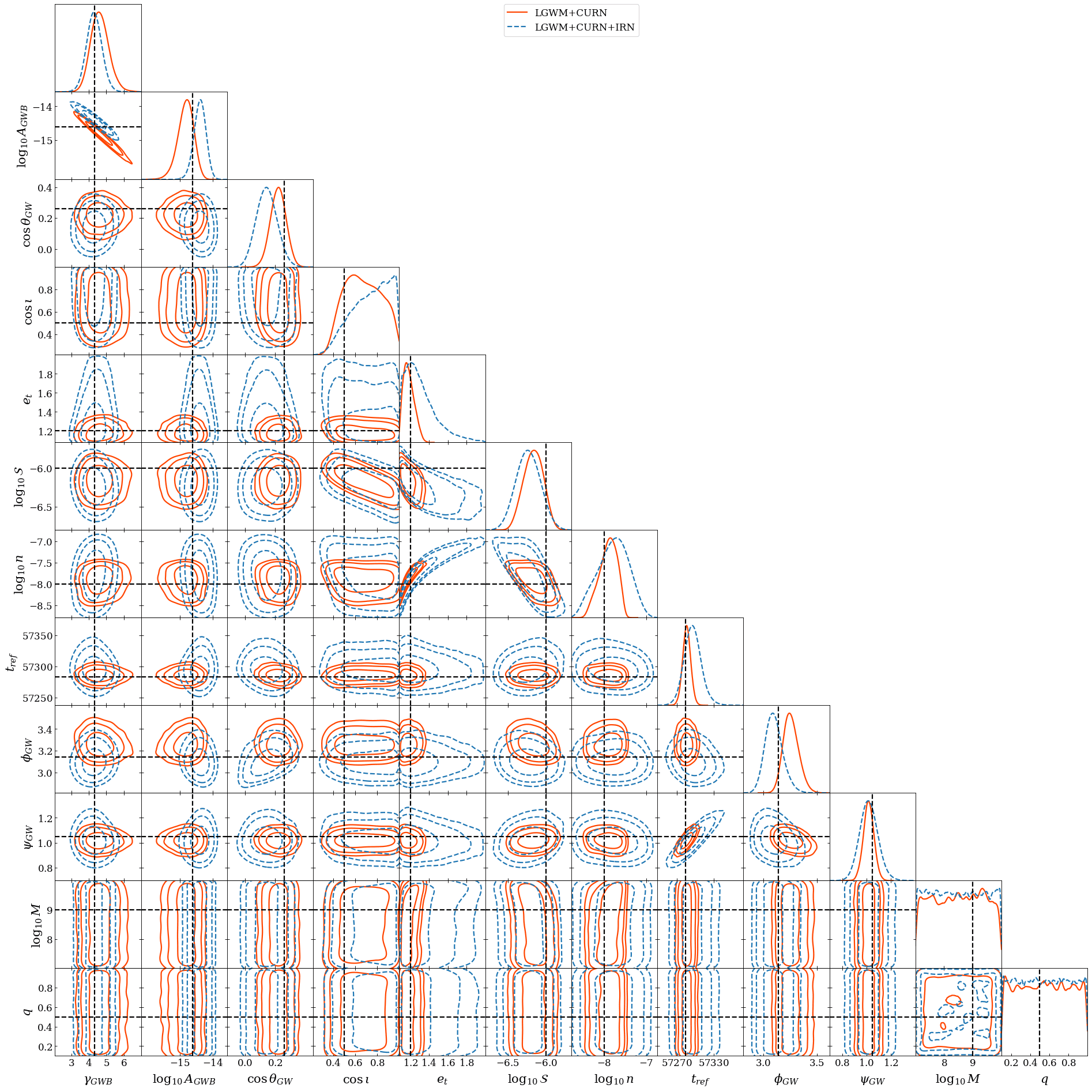}
\caption{
Comparing posteriors that arise from our 
LGWM+CURN and LGWM+IRN+CURN  searches as detailed in Sec.~\ref{simu:44psr} on a simulated NG12.5 dataset containing 44 Pulsars. 
It should be clear that the 
inclusion of IRN parameters does not substantially vary our search results.
These two distinct searches employ the \ptm.}
\label{inj_NG12.5_44psr}
\end{figure*}

 In these simulated datasets, we conduct LGWM+CURN and LGWM+CURN+IRN searches respectively utilizing our pipeline outlined in Section~\ref{sec:LGWM_signals} and employing the priors specified in Table~\ref{param:tab}, specifically the uniform prior on $\log_{10}\mathcal{S}$. For both searches, we utilized \ptm~using various proposals specific to the parameters, such that: \texttt{JumpProposal.draw\_from\_prior} with a weight of 5, and \texttt{JumpProposal.draw\_from\_par\_prior} with a weight of 10 for both CURN and IRN parameters, and 30 for the LGWM parameter. Additionally, we organized three distinct groups, each containing IRN parameters (1 copy), CURN parameters (10 copies), and LGWM parameters (2 copies), which were then passed to \ptm~to enhance convergence. Furthermore, we employed parallel tempering with four geometrically spaced temperatures.
The posterior samples from these two searches are shown as corner plots in Fig.~\ref{inj_NG12.5_44psr}, both for LGWM+CURN and LGWM+CURN+IRN Analyses.

A close inspection of our several corner plots reveals that most of the extrinsic linear GW memory parameters are accurately recovered within a 3$\sigma$ range of the injected parameters.
However, certain intrinsic parameters like the total mass ($M$) and the mass ratio ($q$) of the black hole pairs are rather unconstrained. 
These occurrences are clarified by the given argument. The parameter $M$ influences the PTA signal $s_{+,\times}(t)$ given in Eq.~\eqref{eq:ana} through the amplitude $\mathcal{S}$, regarded as a free parameter, while $q$ affects the signal shape weakly via PN corrections to the orbital dynamics.

From our detailed studies, we gather that it is reasonable to provide an upper bound on $e_t$'s prior at 2.
This is mainly because as $e_t$ increases it would be difficult for the event's PTA signal to remain confined within the entire data span, and therefore will not be detectable.

\begin{figure*}
\centering  
\includegraphics[width=1\linewidth]{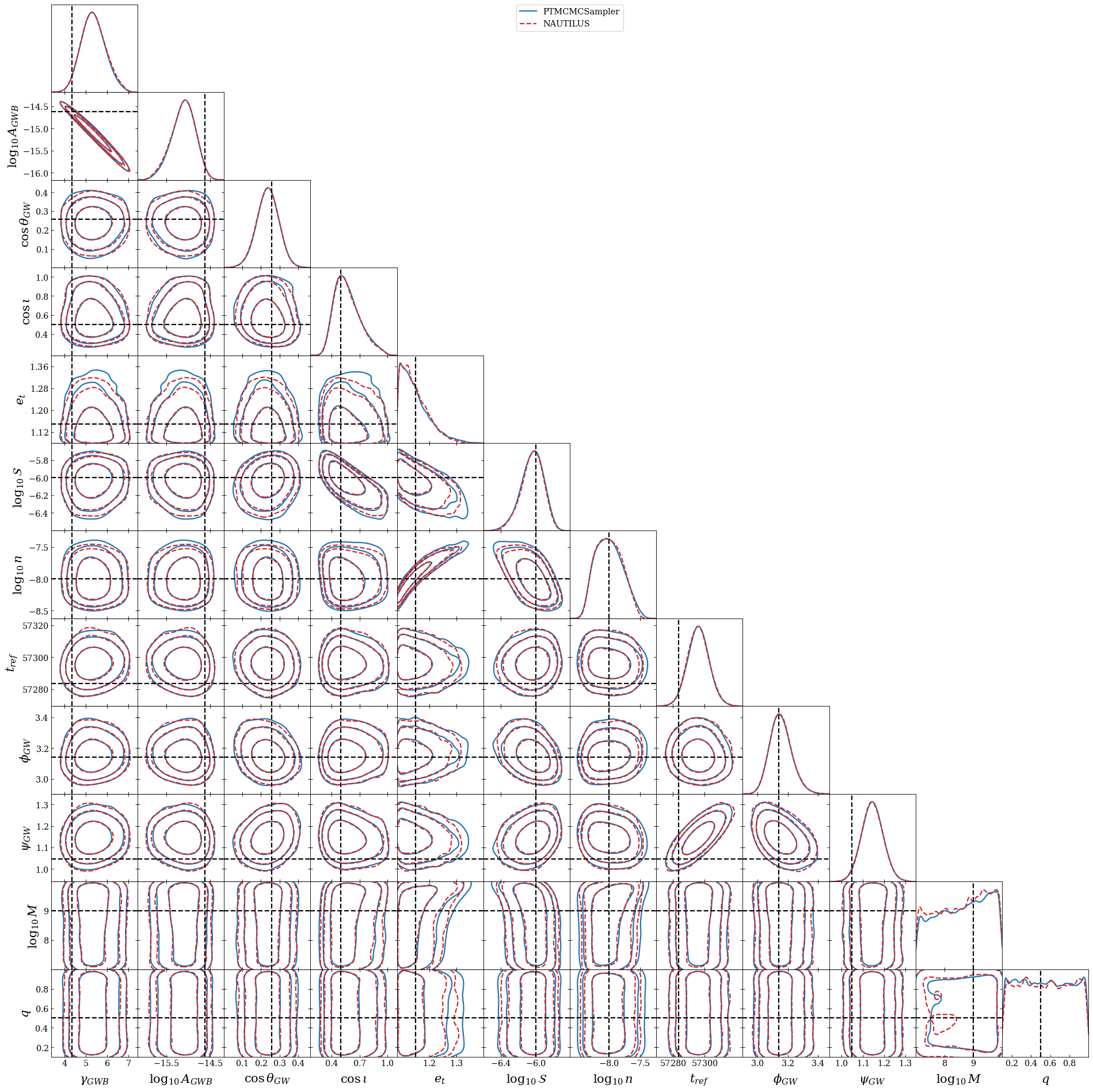}
\caption{
The figure illustrates the posterior distributions resulting from our injection studies involving the LGWM+CURN event on a simulated NG12.5 dataset comprising 44 Pulsars. The covariance among various parameters is evident from the plot. During the simulation, we employed two samplers, \ptm~and \texttt{NAUTILUS}, and observed consistency between the traditional \ptm~typically utilized for PTA searches and one of the most computationally efficient \texttt{NAUTILUS} sampler.
}
\label{inj_NG12.5_44psr_compare_NP}
\end{figure*}

Interestingly, varying $e_t$ values lead to other implications; for example, we infer that the recovery of the inclination angle exhibits a bimodal behavior when we vary $e_t$ values. 
This is evident from a close inspection of Fig.~\ref{inj_NG12.5_44psr_compare_NP} and Fig.~\ref{inj_NG12.5_44psr}.
It should be evident that $e_t=1.15$ run (Fig.~\ref{inj_NG12.5_44psr_compare_NP}) leads to a uni-modal recovery of the orbital inclination ($\iota$) while $e_t=1.2$ (Fig.~\ref{inj_NG12.5_44psr}) run leads to posteriors where $\iota$ exhibits a possibly bimodal distribution.

Furthermore the Figs.~\ref{inj_NG12.5_44psr_compare_NP} and \ref{inj_NG12.5_44psr} clearly demonstrate the correlated nature of various parameter pairs, such as $(\mathcal{S},n), (e_t,n), (\mathcal{S},e_t), (\psi_\text{GW}, t_\text{ref})$, and $(\phi_\text{GW},\psi_\text{GW})$.
These correlations may be attributed to degeneracies that exist in their underlying $s_{+,\times}$.  We now explore the implications of our ready-to-use PTA signal for a realistic PTA dataset.

\subsection{Search on the real NANOGrav 12.5-year Dataset}

\begin{figure*}
\centering  
\includegraphics[width=1\linewidth]{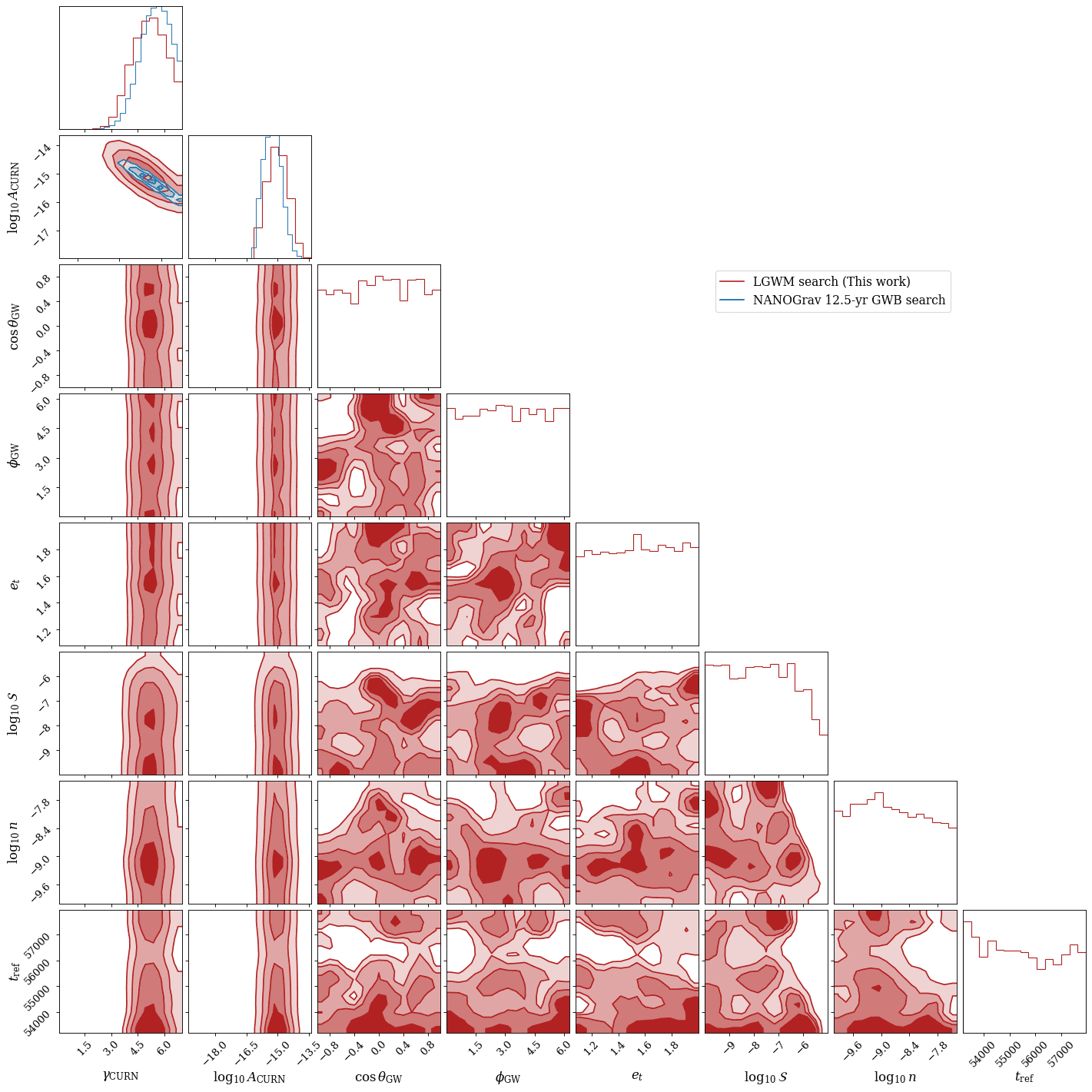}
\caption{ Posterior distributions associated with our LGWM+CURN searches on the real NG12.5 dataset containing the same set of Pulsars as in 
Fig.~\ref{inj_NG12.5_44psr}, namely 44, while 
employing the ~\ptm.
 Clearly, these plots do not show any evidence for the detection of LGWM signal.
The units of each parameter are listed in
table~\ref{param:tab}.
}
\label{real_NG12.5_44psr}
\end{figure*}

We searched for linear GW memory events in the actual NANOGrav 12.5-year (NG12.5) dataset~\cite{agazie2023nanograv} that contains 44 pulsars while employing the \ptm. 
The corner plot in Fig.~\ref{real_NG12.5_44psr} illustrates the marginalized posterior distribution of the linear GW memory event parameters, after accounting for the noise parameters, as obtained from a detection analysis that utilizes a uniform prior on $\log_{10}\mathcal{S}$. 
Clearly, 
Fig.~\ref{real_NG12.5_44psr} does not provide any indications for the presence of our linear GW memory signal. The Bayes factor, which measures the strength of evidence between models that include or exclude the LGWM signal, has been estimated using the Savage-Dickey formula, and it yields a value of approximately~0.68 $\pm0.13$. This value does not support the detection of a LGWM signal in the NG12.5 dataset
(we point to  Ref.~\cite{aggarwal2019nanograv} 
for a detailed discussion of the Savage-Dickey Bayes factor 
computation).

\par
\begin{figure*}[t!]
    \centering
    \includegraphics[width=0.9\textwidth]{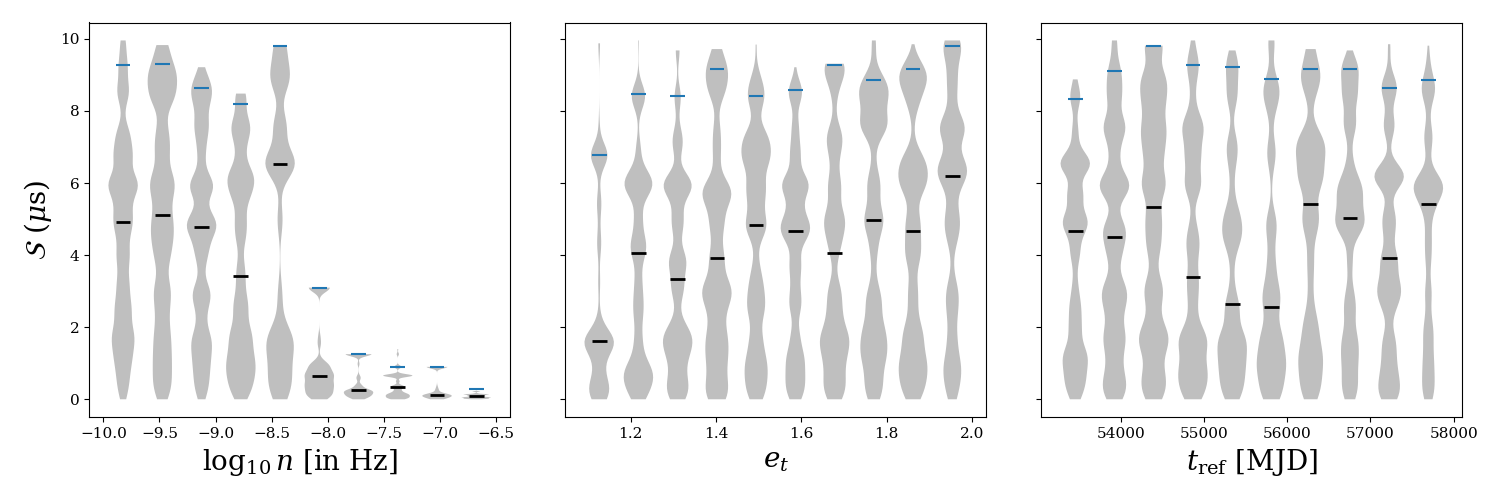} 
      \caption{
      Violin Plot representations 
      for the LGWM amplitude 
      $S$, binned in  $\log_{10} n$ (left plot), $e_t$ (middle plot), $t_\text{ref}$ (right plot) 
      while marginalized over other parameters that 
      are displayed in Fig.~\ref{real_NG12.5_44psr}. The blue horizontal ticks represent the $95\%$ credible upper limits on $\mathcal{S}$ within each category, and the black ticks denote the corresponding medians.}
       \label{violinplot_inj_NG12.5_44psr}
\end{figure*}

 Additionally, we transformed the posterior samples generated from the detection analysis into posterior samples for the upper limit analysis by applying a uniform prior on $\mathcal{S}$, this is accomplished by using a sample reweighting technique, detailed in Ref.~\cite{hourihane2023accurate}. The violin plots of Fig.~\ref{violinplot_inj_NG12.5_44psr} contain three panels that display the posterior distribution for $\mathcal{S}$, grouped into frequency ($n$), eccentricity ($e_t$) and burst epoch ($t_\text{ref}$) bins while considering the marginalization over other parameters. The 95
\% credible upper limits on $\mathcal{S}$ within each specific $\log_{10} n$, $t_\text{tef}$ and $e_t$ bins are shown in Fig~\ref{violinplot_inj_NG12.5_44psr} with blue horizontal ticks. The analysis demonstrates an increase in sensitivity as we shift towards the higher frequency bins, but it remains relatively constant regardless of $e_t$ and $t_\text{tef}$.

\begin{figure}[t!]
\centering
\includegraphics[width=0.45\textwidth]{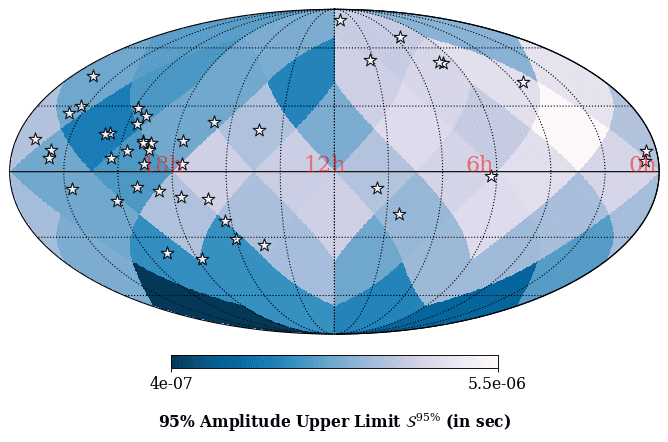} 
  \caption{
   The upper limits on  our linear GW memory amplitude as a function of 
the  pixilated sky position while employing the NG12.5 dataset 
in our LGWM+CURN searches. The constraints are tighter in the hemisphere 
characterized by a higher pulsar density.}
   \label{skyplot_NG12.5_44psr}
\end{figure}

Thereafter, we compute the 95\% upper limit on LGWM signal amplitude ($\mathcal{S}$) while marginalizing over other parameters. Samples with $n > 3.16$ nHz are excluded because are significantly influenced by the prior distribution, rendering them non-informative as discussed in Ref.~\cite{Lokie2023nanograv}. The 95\% upper limits on $\mathcal{S}$ are determined to be 3.48 $\pm.51 \ \mu$s for $n>3.16$ nHz, with uncertainties calculated through the bootstrap method. Additionally, we overlay the posterior distributions for CURN parameters from the NG12.5 GWB search~\cite{arzoumanian2020nanograv} in Fig.~\ref{real_NG12.5_44psr}. It is observed that introducing the LGWM signal does not modify the posterior distribution of CURN parameters in our analyses. This highlights the robustness of our search against potential power leakage between CURN and the linear GW memory signal.
Furthermore,  we compute upper limits on linear GW memory strain amplitude in the NANOGrav 12.5-year dataset, showcasing their dependence on the sky location ($\alpha_\text{GW}$ and $\delta_\text{GW}$) of the GW source 
as displayed in Fig~\ref{skyplot_NG12.5_44psr}. This analysis encompasses the marginalization over all other parameters and incorporates a variable common red-noise spectral index. Utilizing the \texttt{HEALPix} method,
described in Refs.~\cite{zonca2019healpy, gorski2005healpix},
the sky is divided into 48 pixels (\texttt{nside}=2). 
NG12.5 Pulsar locations are marked in the plot, revealing that the upper limit is more precisely constrained in regions with a higher pulsar density.

\begin{figure}
    \centering
    \includegraphics[width=0.5\textwidth]{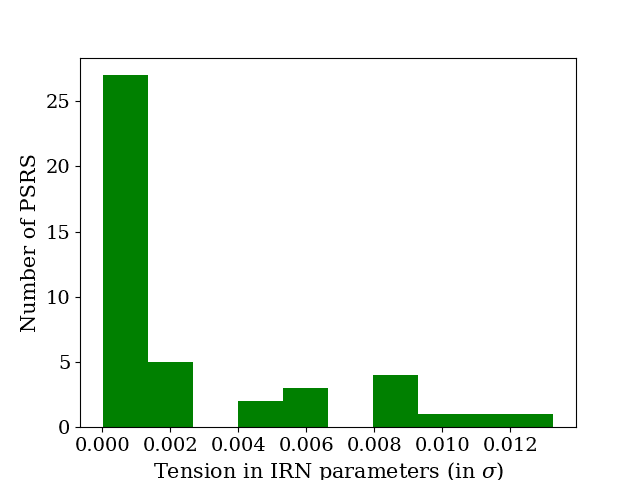} 
      \caption{A histogram  to display the Raveri-Doux tension between the 
 marginalized posterior distributions of IRN parameters that arise from the  NANOGrav 12.5-year GWB search and the present effort.
 The tension stays below 0.014$\sigma$ for all pulsars, indicating that the presence of IRN has a negligible impact on our search results.}
       \label{tension_NG12.5_44psr}
\end{figure}

We now evaluate 
the resilience of our search pipeline to power leakage between the IRN and the hyperbolic LGWM signal. To do this, we compare the posterior distributions of the IRN parameters obtained from LGWM detection analyses with those derived from the NG12.5 GWB search, detailed in
Ref.~\cite{arzoumanian2020nanograv}.
For this comparison, we employ 
the Raveri-Doux tension statistic, which is available in the \texttt{tensiometer} package~\cite{raveri2021non}.
In Fig.~\ref{tension_NG12.5_44psr}, a histogram of \texttt{tensiometer} statistics is presented for various pulsars, and we observe that the IRN parameter posteriors from our analyses closely match those from the NG12.5 GWB search~\cite{arzoumanian2020nanograv} within a 0.015$\sigma$ range for all pulsars. This finding instills 
certain confidence
that our search for the LGWM signal remains largely unaffected by the presence of IRN.

\section{Summary and Discussion}
\label{sum:dis}
We have provided an efficient way to describe linear GW memory (LGWM) events in PTA datasets. These events are due to  GWs from hyperbolic passages of SMBHs 
and we employed the PN approach to describe the resulting temporally evolving GW polarisation states. 
The resulting timing residuals involve 
burst and ramp functions. Further, we pursued two types of simulation studies. In the first set, we introduced CURN and our GW memory signal (LGWM+CURN). In the second set, we introduced an additional component, namely IRN(LGWM+CURN+IRN). 
In both scenarios, we observed that the marginalized posterior distributions of our free model parameters closely matched the injected values for the simulated data.
Subsequently, we conducted an all-sky search for LGWM events in the NANOGrav 12.5-year dataset. Our model includes a CURN process, identified as a precursor to the possible GWB in the NG12.5 dataset~\cite{arzoumanian2020nanograv}, along with IRN and WN processes for each pulsar, in addition to our LGWM signal. 
Further, we kept the WN parameters fixed for each pulsar based on values obtained from its single-pulsar noise analysis. This marked the first instance of a search for GWs arising from hyperbolic encounters of Black Holes conducted on a full-scale PTA dataset.

\par Despite the comprehensive search, we did not discover any evidence of LGWM signals in the  NG12.5 dataset.
 Consequently, we computed the upper limits on the GW signal amplitude ($\mathcal{S}$) of the SMBHB candidate, taking into account the eccentricity ($e_t$), characteristic frequency ($n$), and burst epoch ($t_\text{ref}$). The 95\% upper limits on LGWM amplitude ($\mathcal{S}$), marginalized over all other parameters, are 3.48 $\pm.51 \ \mu$s for $n>3.16$ nHz.
Naturally, the present effort should allow us to pursue a detailed search for such events in the IPTA DR3 dataset in the coming years.

 There are several possible directions that we plan to pursue 
 soon.
It will be interesting to 
explore if globular cluster pulsars should be ideal for detecting such LGWM events. It is reasonable to expect that 
pulsar terms could be relevant for such a scenario where our LGWM events can appear in the timing data of globular cluster MSPs with certain time 
offsets.  It should be noted that such pulsars are not included in the current PTA efforts due to various difficulties in pursuing deterministic pulsar timing of globular cluster pulsars, attributed to the stochastic nature of globular cluster 
gravitational potential \cite{blandford1987globular}.
It may be recalled that we  
we neglected the effects of pulsar terms in the present effort. This is justified as PTA pulsars are generally situated hundreds to thousands of light years away from the SSB, resulting in transient durations much shorter than the time it takes for the pulses to reach the SSB.

It will be interesting to include spin effects into the current PN
accurate description of non-spinning BHs in hyperbolic orbits and explore 
its PTA implications
\cite{de2014gravitational}.
Another potential direction should be to employ 
the Effective-One-Body approach for describing the dynamics of 
unbound BHs \cite{damour2016effective}.
This approach should be appropriate for describing $h_{\times,+}(t)$ 
associated with the part of the parameter space that we neglected 
in the present effort~\cite{nagar2021effective}.
Recall that we imposed such restrictions 
to ensure that 
 the PN approximation is indeed valid. 
 Another direction involves exploring possible degeneracy between our prescription and the burst signal due to cosmic strings \cite{yonemaru2021searching}.
 It may be noted that such burst events arise through cusps and kinks 
 associated with string re connection events \cite{damour2001gravitational}. 
 From a computational perspective, it will be valuable to extend 
 the \texttt{QuickCW}~\citep{becsy2022fast} method to 
 include 
 our linear GW memory searches as this method  
 leverages the mathematical structure of PTA signal expressions to expedite likelihood computations during projection parameter updates (as outlined in Table~\ref{param:tab} in our case).
 This extension should be essential to maintain computational feasibility in upcoming IPTA searches given the anticipated growth in data volumes. Another novel direction involves including the 
 Hellings-Downs correlations instead of the current CURN SGWB signal. It should be possible re-weight our CURN to achieve the 
 Hellings-Downs
 correlations by adapting  the method described in~\cite{hourihane2023accurate}. Further, it should be worthwhile to explore possible astrophysical rates for 
 our linear memory events.

\section*{Software}
\epr\cite{ellis_2020_4059815,johnson2023nanograv}, \texttt{enterprise\_entensions}~\cite{TaylorBaker+2021, johnson2023nanograv}, \texttt{PTMCMCSampler}~\cite{ellis2017jellis18}, \nau\cite{lange2023nautilus}, \texttt{GW\_hyp}~\cite{dandapat2023gravitational}, 
\texttt{GWecc.jl}~\cite{susobhanan2020pulsar,susobhanan2023eccentric},
\texttt{libstempo}~\cite{vallisneri2020libstempo}, 
\texttt{numpy}~\cite{HarrisMillman+2020}, 
\texttt{wquantiles}~\citep{Sabater2015},  
\texttt{tensiometer}~\citep{raveri2021non},
\texttt{matplotlib}~\citep{Hunter2007}, 
\texttt{corner}~\citep{Foreman-Mackey2016},
\texttt{healpy}~\cite{zonca2019healpy, gorski2005healpix}.

\section*{Acknowledgements}
We thank Johannes U. Lange, Aurélien Chalumeau, Shubhanshu Tiwari, Golam Shaifullah, Chan Park, and Muyoung Heo for helpful discussions and the unknown referee for valuable 
suggestions.
SD and AG acknowledge the support of the Department of Atomic Energy, Government of India, under project identification \# RTI 4002.
AS is supported by the NANOGrav NSF Physics Frontiers Center (awards \#1430284 and 2020265). 
LD is supported by a West Virginia University postdoctoral fellowship and is a member of the NANOGrav PFC supported by NSF award \#2020265.
Computation for this research was supported by the Space High-performance Computing Center at the Institute for Basic Science.

\begin{widetext}
\appendix
\counterwithin{figure}{section}
\section{Comparing with \gec\ Package and Limitations in Achieving Full Recovery for \gh\ with Single Pulsar}
\label{app:gwecc:gwhyp}
In this section, we will compare the effectiveness of recovering parameters using the \gh\ package to the \gec~\footnote{\url{https://github.com/abhisrkckl/GWecc.jl}} package, as described in Ref.~\citep{susobhanan2023eccentric,susobhanan2020pulsar}. It is important to consider that eccentric signals consist of multiple cycles, while hyperbolic encounters only have a single cycle, often with a bump at the time of encounters. Additionally, the memory term indicates a difference in slope between the beginning and end of the residuals with respect to time. Since LGWM events contain only one cycle, we should not expect to achieve full recovery for hyperbolic searches compared to eccentric searches. On the other hand, when an eccentric signal contains only a portion of the cycle, it behaves similarly to LGWM residuals. We conducted two parameter estimation studies:
\begin{enumerate}
    \item We focused on a part of the cycle of eccentric signals and searched for eccentric signals in simulated data.
    \item We consider multiple cycles of eccentric signals within the same data span and identify these eccentric signals in simulated data.
\end{enumerate}
Our findings showed that for case 1, the recovery of true parameters was significantly beyond 3$\sigma$, while, in case 2, we achieved excellent recovery within 1$\sigma$, as illustrated in Fig.~\ref{1Ncyc}. This result was expected because in case 1, we only considered one cycle, which explains why proper recovery for LGWM search with a single pulsar was not expected. However, incorporating multiple pulsars improved the recovery, as we observed in our previous analyses. 

\begin{figure*}[t!]
\centering
\includegraphics[width=0.9\textwidth]{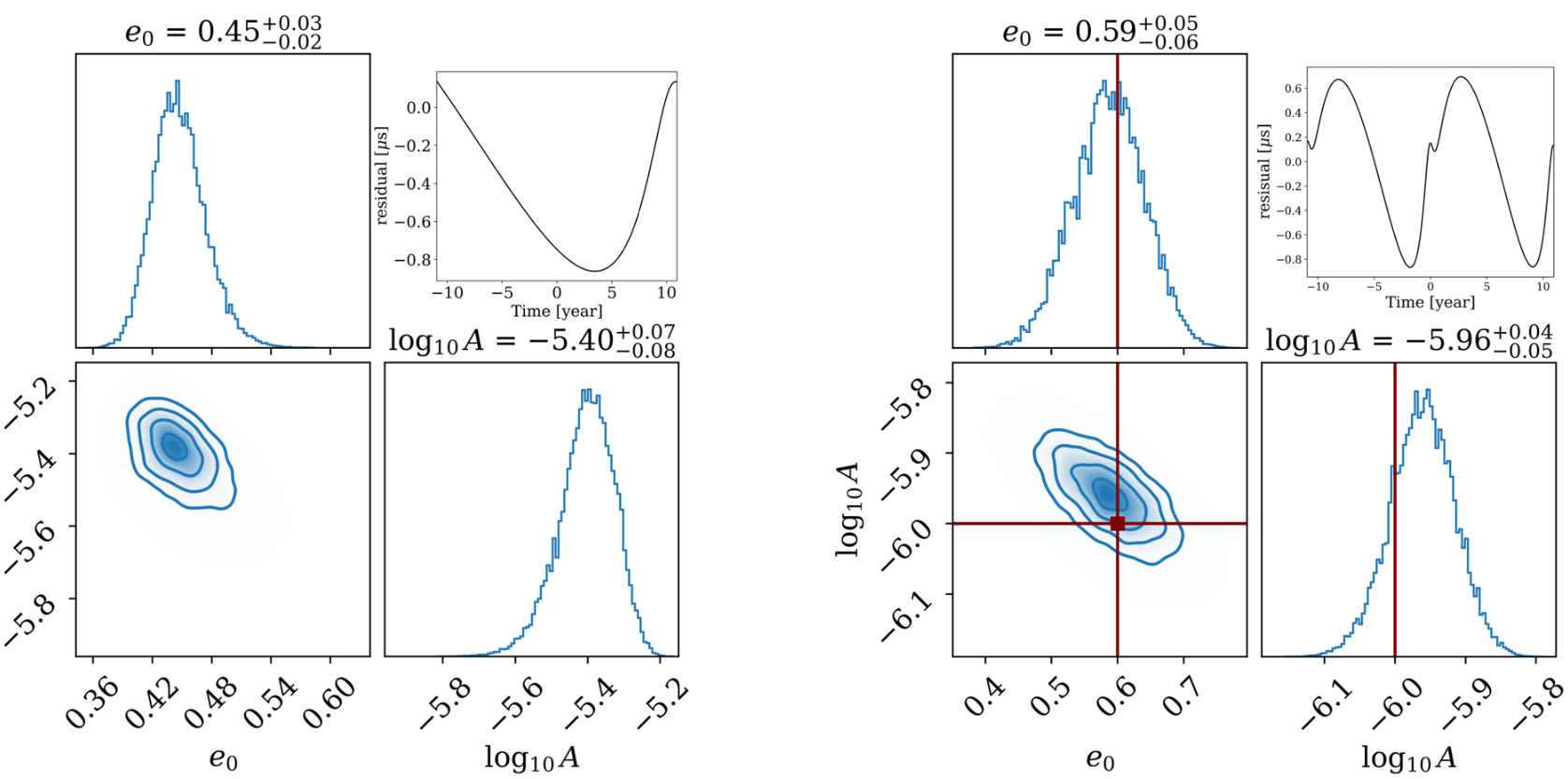}  
\caption{Parameter estimation (PE) using \gec. The left plot shows single-cycle PE, while the right plot displays multi-cycle PE on a simulated dataset. The residual plot is provided in the top right corner of each of these plots.}
\label{1Ncyc}
\end{figure*}

\end{widetext}

\bibliographystyle{apsrev4-1}
\bibliography{mybib}

\end{document}